\documentclass[preprint,12pt,authoryear]{elsarticle}

\usepackage{amssymb}
\usepackage{amsmath}
\usepackage{lineno}
\usepackage{siunitx}
\usepackage{float}
\usepackage{booktabs}
\usepackage{multirow}
\usepackage{adjustbox}
\usepackage{threeparttable}
\usepackage{longtable}
\usepackage{placeins}

\journal{Earth and Planetary Science Letter}

\begin{document}

\begin{frontmatter}



\title{A surviving pink spinel records an early aluminous melt on the ureilite parent body} 

\author[1,2,3]{Yaozhu Li}
\author[2,3]{Phil J. A. McCausland}
\author[2,3]{Roberta L. Flemming}
\author[4]{Noriko Kita}
\author[1]{Carsten Detlefs}

\affiliation[1]{organization={ID03, European Synchrotron Radiation Facility},
               addressline={71 Avenue des Martyrs},
               city={Grenoble},
               postcode={38000},
               country={France}}

\affiliation[2]{organization={Department of Earth Sciences, University of Western Ontario},
               addressline={1151 Richmond Street},
               city={London},
               state={Ontario},
               postcode={N6A 5B7},
               country={Canada}}

\affiliation[3]{organization={Institute for Earth and Planetary Exploration, University of Western Ontario},
               addressline={1151 Richmond Street},
               city={London},
               state={Ontario},
               postcode={N6A 5B7},
               country={Canada}}

\affiliation[4]{organization={Department of Geoscience, University of Wisconsin--Madison},
               addressline={1215 West Dayton Street},
               city={Madison},
               state={WI},
               postcode={53706},
               country={United States}}

\begin{abstract}
Ureilites are ultramafic achondrite meteorites that likely represent fragments of a differentiated parent body. However, the origin and evolution of the ureilite parent body remains debated, as textural equilibrium is commonly observed alongside chemically primitive compositions. Conventionally, ureilites are interpreted as mantle residues from a large rocky body, where the scarcity of feldspathic materials has been attributed to early accretion and extensive partial melting. Here we report new mineralogical, isotopic, and microstructural observations from polymict ureilite breccia Elephant Moraine (EET) 87720. The ureilite sample contains unusually magnesian olivine clasts with compositions reaching Mg\# = 98.7, together with calcium-poor pyroxene (Wo as low as 1.0). We also identify coarse-grained aluminous pink spinel clasts containing 56.4--58.7 wt\% Al$_2$O$_3$ and 11.3--11.8 wt\% Cr$_2$O$_3$, which are rare among ureilites. In situ triple oxygen isotope measurements of aluminous spinel and associated forsteritic olivine using Secondary Ion Mass Spectrometry plot along the $\sim$1 slope Carbonaceous Chondrite Anhydrous Mineral (CCAM) line, consistent with bulk ureilites, including samples from Almahata Sitta. These clasts also follow the Fe-loss/addition trend defined by molar FeO/MnO versus FeO/MgO relationships, showing near-constant chondritic Mn/Mg ratios. Together, these observations demonstrate that the clasts are indigenous to the ureilite parent body and extend the known oxygen isotope range of ureilites to $\delta^{18}$O $\sim$ 9.7‰. Three-dimensional dark-field X-ray microscopy (DFXM) reveals a hierarchical deformation microstructure within the aluminous spinel, comprising distributed lattice curvature, localized slip-band-like boundaries, and coherent mosaic-domain boundaries. These observations demonstrate that shock deformation was accommodated through multiple scales of crystallographic lattice subdivision and provide the first three-dimensional characterization of crystal-plastic deformation preserved within an aluminous spinel from the ureilite parent body. We propose a magmatic origin for the aluminous spinel through crystallization from a locally Al-rich, Ca-poor melt, likely under low oxygen fugacity, favouring Al-rich spinel over Cr-rich endmembers. Using Al partitioning between coexisting spinel and olivine, we estimate a crystallization temperature of $1318 \pm 43$ K. This temperature agrees well with previous estimates for ureilites and supports crystallization within a thermally elevated ureilite parent body during the early evolution of the Solar System. The spinel may therefore preserve a crystallization product of an early aluminous melt that has largely disappeared from the preserved ureilite record, providing a complementary archive of the earliest stages of planetary differentiation.

\end{abstract}


\begin{highlights}
\item Oxygen isotopes confirm aluminous pink spinel is indigenous to the ureilite parent body
\item Triple oxygen isotope data extend the ureilite range to $\delta^{18}$O $\sim$ 9.7‰
\item DFXM reveals three level hierarchy of microstructure within the aluminous spinel
\item The preservation of Al-spinel suggests a very early melt in UPB that largely disappeared in the main-group of ureilites 
\end{highlights}

\begin{keyword}
Ureilites, Spinel, Triple-oxygen isotopes, DFXM
\end{keyword}
\end{frontmatter}



\section{Introduction}
Ureilites are ultramafic achondrite meteorites composed primarily of olivine and pyroxene, yet they retain primitive chondrite-like characteristics, including high carbon abundance, enrichment in siderophile elements, and elevated noble gas contents \citep{Goodrich1992, GoodrichEtAl2004, GoodrichEtAl2015, LiEtAl2021, Rubin2006}. The parent body of ureilites remains uncertain; however, these meteorites are widely interpreted to originate from a sizable rocky body that experienced partial differentiation \citep{GoodrichEtAl2004, GoodrichEtAl2015, WilsonEtAl2008}. Reconciling the coexistence of textural equilibrium with chemically primitive compositions remains a central challenge in understanding the formation and evolution of the ureilite parent body.

Almost all ureilite samples are brecciated and shocked, and they are commonly classified into monomict and polymict ureilites based on mineralogy and texture. Monomict ureilites constitute approximately 90\% of known ureilite meteorites and typically display coarse-grained olivine and pigeonite (low calcium pyroxene, Wo 5–14) forming triple-junction grain boundaries, indicative of textural equilibrium. The magnesian content (Mg\#, defined as Mg/(Mg+Fe), commonly in range ~Fo 75 to Fo 95) varies significantly among different monomict samples but is generally uniform within a single specimen. A decreasing trend in calcium content in pyroxene with increasing Mg\# in olivine has been documented \citep{GoodrichEtAl2004, GoodrichEtAl2015, WilsonEtAl2008}. The olivine--pigeonite assemblage is the most common mineral association in monomict ureilites, whereas olivine--augite and olivine--orthopyroxene assemblages are comparatively rare \citep{GoodrichEtAl2004, GoodrichEtAl2015, WilsonEtAl2008}. 

Compositional relationships between olivine and pyroxene, expressed as FeO/MnO versus FeO/MgO ratios, commonly display near-chondritic trends, suggesting a primitive origin as partial melt residues \citep{GoodrichEtAl2004}. Similarly, oxygen isotope compositions (e.g., $\delta^{18}$O versus $\delta^{17}$O) for ureilites typically plot along the Carbonaceous Chondrite Anhydrous Mineral (CCAM) line, reflecting a chondritic affinity \citep{DownesEtAl2008, GoodrichEtAl2004}. However, felsic minerals such as feldspar or silica are extremely rare in ureilites, and their scarcity has been attributed to extensive partial melting, estimated to have removed approximately 15--25\% of the original material \citep{GoodrichEtAl2015, WilsonEtAl2008}. Consequently, ureilites are commonly interpreted as partial melt residues or \textit{paracumulates}, a term introduced by \citet{WarrenKallemeyn1989} to describe materials that preserve both residual and cumulate characteristics.

Approximately 10\% of ureilites are polymict ureilites, which are texturally and mineralogically distinct from monomict ureilites in that they contain large angular lithic and mineral clasts embedded within a fine-grained matrix. These clasts may be either exogenous or endogenous to the ureilite parent body \citep{Goodrich1992, WarrenKallemeyn1989}. Polymict ureilites therefore provide access to a broader range of lithologies from the parent body and may preserve critical information on its internal structure and evolutionary history. For example, basaltic components containing approximately 2 vol\% feldspathic material have been reported in polymict ureilites such as Dar al Gani (DaG) 164/16, DaG 319, DaG 665, and Elephant Moraine (EET) 83309 \citep{CohenEtAl2004}. 

Some Almahata Sitta (AhS) breccias are considered anomalous polymict ureilites because they contain a diverse assemblage of ureilitic materials, including monomict ureilite clasts and rare andesitic components of ureilitic origin \citep{GoodrichEtAl2015, HorstmannBischoff2014}. In addition, AhS breccias contain materials derived from other primitive asteroid bodies, including enstatite, ordinary, carbonaceous, and Rumuruti-like meteorites. These observations have led to the hypothesis that a ureilite daughter body formed as a rubble-pile asteroid composed of mixed asteroidal materials following catastrophic disruption of the original ureilite parent body \citep{GoodrichEtAl2015, HorstmannBischoff2014}.

 The polymict ureilite EET 87720 has previously been studied for its mineralogy, texture, bulk composition, oxygen isotopes, and shock history \citep{DownesEtAl2008, GoodrichEtAl2004, LiEtAl2021, RaiEtAl2003, WarrenKallemeyn1992}. In particular, \citet{LiEtAl2021} estimated a shock stage of at least S5 and first documented the presence of an abnormally large Al-rich spinel occurring as a discrete clast rather than as an interstitial phase. Spinel commonly crystallizes in equilibrium with forsteritic olivine and has long been used as a geothermometer to estimate crystallization conditions \citep{RoederEtAl1979, WanEtAl2008}. The discovery of this aluminous spinel therefore provides a unique opportunity to investigate the lithology and thermal history of the ureilite parent or daughter body.

To date, spinel group oxides are rarely documented in ureilites, and when present, they are typically chromite. For example, \citet{ChikamiEtAl1997} reported chromium-rich spinel with compositions of Cr$_2$O$_3$ 54--56 wt\% and Al$_2$O$_3$ 14--15.6 wt\% in Lewis Cliff (LEW) 88774. Titanium-bearing spinel (ulv\"ospinel) has also been reported in DaG 665 and EET 88309 in association with labradoritic plagioclase (An$_{40-58}$) clasts \citep{CohenEtAl2004}. The occurrence of coarse-grained aluminous spinel in EET 87720 therefore represents an unusual and potentially diagnostic mineralogical feature.

In this study, we investigate the origin and deformation history of the aluminous spinel clast in polymict ureilite EET 87720, specifically the thin section (30 $\mu$m thickness) specimen TS 57, the same thin section studied by Li et al. (2021),  using a combination of oxygen isotope geochemistry, mineral chemistry, thermometry, and three-dimensional X-ray diffraction microscopy. Triple oxygen isotope measurements obtained by secondary ion mass spectrometry (SIMS) are used to evaluate the relationship between the spinel and known ureilitic reservoirs. Equilibrium relationships between spinel and olivine are further used to constrain crystallization conditions and compare them with existing thermal evolution models for the ureilite parent body.

Finally, we examine the internal crystallographic structure of the spinel using dark-field X-ray microscopy (DFXM) at beamline ID03 of the European Synchrotron Radiation Facility (ESRF). DFXM provides non-destructive three-dimensional imaging of lattice orientation and mosaicity at at submicrometer  to nano resolution, allowing the volumetric reconstruction of the internal deformation microstructures in the spinel. The reconstructed 3D volume reveals spatially coherent mosaic domains and localized low-angle subgrain boundaries that persist through depth within the crystal \cite{LiEtAl2026MarsStrain, ZelenikaEtAl2024, IsernEtAl2025, GarrigaEtAl2023}. These observations provide new constraints on the relationship between crystallization, shock deformation, and intracrystalline strain accommodation in refractory phases hosted within EET 87720.

\section{Methods}
\subsection{Electron probe microanalysis}

Electron probe microanalysis (EPMA) was conducted at the University of Western Ontario using a JEOL JXA-8530F field-emission electron microprobe equipped with five wavelength-dispersive spectrometers (WDS). Quantitative analyses of major element oxides were obtained for mineral phase identification and compositional characterization of EET 87720. The elemental X-ray line positions of standards were obtained to ensure reliable results.

Backscattered electron (BSE) mapping was performed using an accelerating voltage of 15 kV and a beam current of 20 nA with a dwell time of 0.2 ms. A total of 36 individual beam scans were acquired and subsequently stitched together using JEOL software to produce a high-resolution BSE mosaic image. Major element oxide abundances, including Si, Al, Na, Mg, K, Ca, Ti, Fe, Mn, Ni, and Cr, were measured to determine mineral compositions and phase relationships.

\subsection{In situ secondary ion mass spectrometry}
Oxygen three-isotope ratios were obtained by non-destructive in situ secondary ion mass spectrometry (SIMS) using the IMS 1280 instrument at the WiscSIMS Laboratory, following analytical procedures similar to those described by \citet{kita2010high} and \citet{zhang2022sims}. A 2 nA Cs\(^+\) primary ion beam focused to approximately 12 \(\mu\)m produced \(^{16}\mathrm{O}^{-}\) intensities of \(\sim 2.4 \times 10^{9}\) cps.

Oxygen isotopes \(^{16}\mathrm{O}^{-}\), \(^{17}\mathrm{O}^{-}\), and \(^{18}\mathrm{O}^{-}\) were measured simultaneously using multicollection Faraday cups equipped with feedback resistors of \(10^{10}\), \(10^{12}\), and \(10^{11}\ \Omega\), respectively. Mass resolving powers were approximately 2200 for \(^{16}\mathrm{O}^{-}\) and \(^{18}\mathrm{O}^{-}\), and approximately 5000 for \(^{17}\mathrm{O}^{-}\). Each analysis required approximately 7 minutes, including presputtering, secondary ion centering, and signal acquisition (\(\sim 200\) s).

At the conclusion of each analysis, \(^{16}\mathrm{O}^{1}\mathrm{H}^{-}\) intensities were monitored to evaluate potential tailing corrections on \(^{17}\mathrm{O}^{-}\). Measured contributions (\(8 \pm 2\) ppm of \(^{16}\mathrm{O}^{1}\mathrm{H}^{-}\)) were negligibly small (\(<0.1\)\textperthousand). Typically, approximately 15 unknown analyses were bracketed by four sets of San Carlos olivine standard analyses (SC-Ol; \citealt{kita2010high}) to correct for instrumental mass fractionation.

A suite of matrix-matched standards consisting of one spinel, two chromite, five olivine, and five pyroxene standards \citep{heck2010single, zhang2022sims} was analyzed during the same analytical session to determine instrumental bias corrections as a function of molar Al/(Al+Cr) in spinel, Fo content in olivine, and En--Fs--Wo composition in pyroxene. Typical spot-to-spot reproducibility of the SC-Ol standard analyses was 0.12\textperthousand\ for \(\delta^{18}\mathrm{O}\) and 0.25\textperthousand\ for both \(\delta^{17}\mathrm{O}\) and \(\Delta^{17}\mathrm{O}\), and these values were adopted as uncertainties for individual unknown analyses.

\subsection{Dark-field X-ray microscopy}
\begin{figure}
    \centering
    \includegraphics[width=1\linewidth]{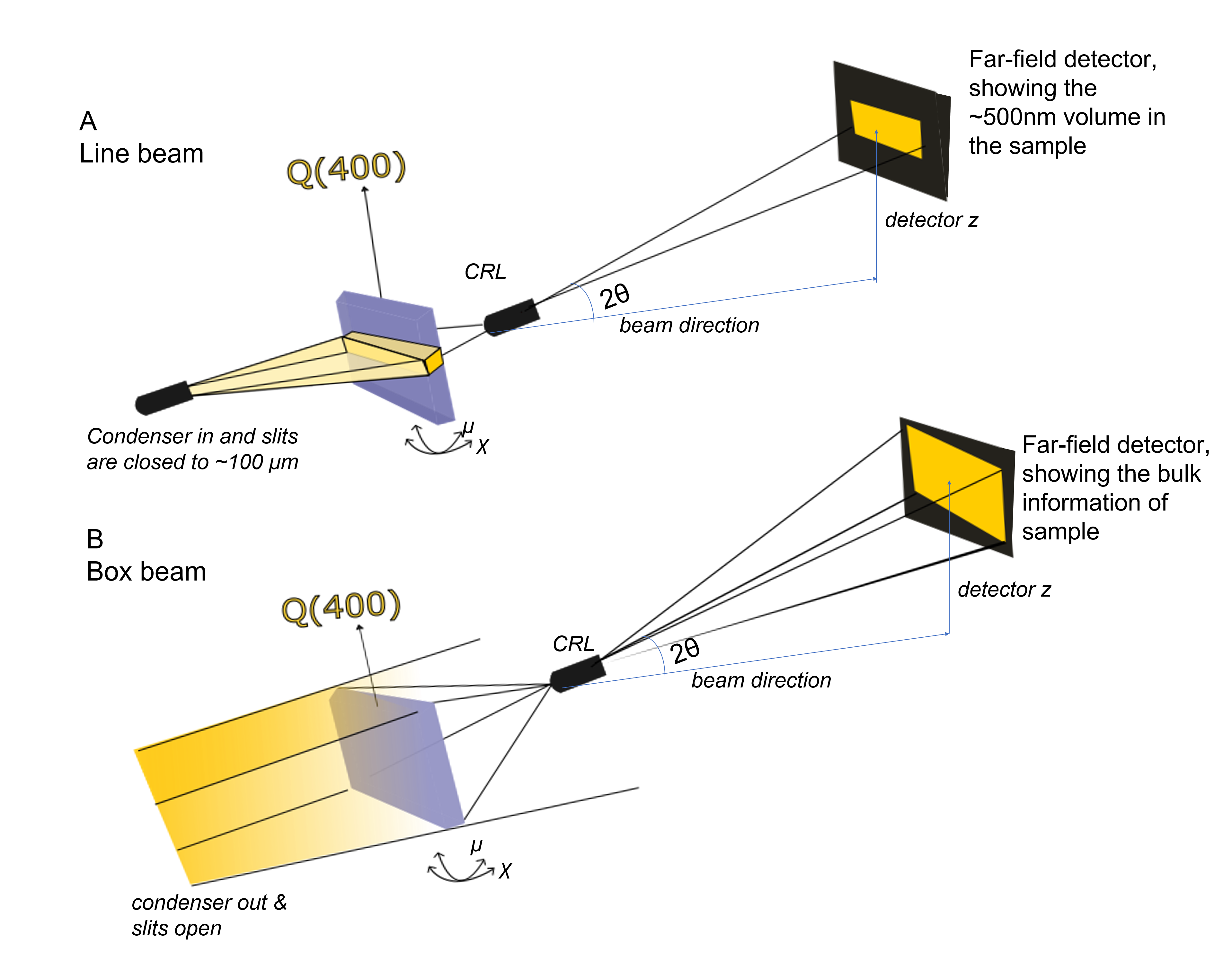}
    \caption{Illustration of DFXM geometry. \textbf{A} shows a line beam configuration. Condenser is in and slits are usually closed to 100 $\mu$m for the best line beam. Compound-refractive lens (CRL) objective module is tilted to accordingly to the 2$\theta$ angle. Far-field detector is moved up to capture the diffraction image. \textbf{B}shows a box beam configuration. Condenser is out and slits are open. In such configuration, the bulk sample information are captured in the far field detector
    }
    \label{dfxm}
\end{figure}

The Dark-Field X-Ray Microscopy (DFXM) experiments were performed at the ESRF beamline ID03 \citep{IsernEtAl2025}. DFXM is a diffraction imaging technique where a crystalline element within the sample (e.g.~a grain) is aligned to the desired Bragg reflection.
An X-ray objective lens is then placed into the Bragg diffracted beam between the sample and a high-resolution imaging detector, allowing the capture of the real-space image of defects (dislocations, stacking faults, strain fields, \ldots) in the crystal lattice \citep{SimonsEtAl2015, PoulsenEtAl2017, PoulsenEtAl2021}. The data for EET 87720 were collected using photon energy \SI{17}{keV}, selected by a Double Multilayer Monochromator (DMM) followed by a Channel-cut Crystal Monochromator (CCM, Si(111) ) with a bandwidth of $\Delta E/ E \approx 10^{-2}$ and $\Delta E/ E \approx 10^{-4}$, with stored current of \SI{198}{mA}. We selected lattice plane (400) for this study, given the available working range of the detectors and sample goniometer. During experiment, to facilitate the 3D reconstruction, the monochromatic beam was focused in the vertical direction by a Compound Refractive Lens (CRL) condenser comprised of 58 bi-parabolic (1D focusing) Be lenslets with an $R=\SI{100}{\um}$ radius of curvature, yielding an effective focal length of \SI{908}{mm}. The beam profile on the sample was approximately $\SI{200}{\um}(h) \times \SI{0.6}{\um}(v)$ (FWHM). The horizontal \textit{line beam} illuminated a single plane that sliced through the depth of the crystal, defining the microscope's \textit{gauge volume} or ``layer'' (Fig. ~\ref{dfxm}A). We then traversed the sample through the line beam, enabling the scan of different volumes in the sample and further facilitating a 3D dataset by stacking layers accordingly. In contrast, a \textit{box beam} is the beam in which the slits are open up to \SI{0.5}{mm} * \SI{0.5}{mm} without using the CRL condenser (Fig. ~\ref{dfxm}B). In this configuration, the diffraction image is called a ``projection'' of the sample,  providing overall diffraction information of the whole illuminated volume in the sample.

The far-field imaging detector used an indirect X-ray detection scheme. This detector was comprised of a scintillator crystal to convert X-rays into visible light, a visible light microscope and sCMOS camera (PCO.edge 4.2bi with $2048 \times 2048$ pixels). It was positioned \SI{5010}{mm} downstream of the sample. The visible light optics inside the far-field detector could switch between $10\times$ and $2\times$ magnification to achieve an effective pixel size of \SI{0.65}{\um} or \SI{3.25}{\um} at the scintillator, and \SI{36.3}{nm} or \SI{182}{nm} at the sample position (Fig. ~\ref{dfxm}), respectively, and an angular resolution of \SI{0.1}{millidegree}. The data was processed by \textit{darfix} \citep{GarrigaEtAl2023}, similar to the previous work done by \citet{LiEtAl2026MarsStrain}.

In this work, we collected \textit{mosaicity map} by rotating two orthogonal motors, $\mu -\chi$ respectively, in the lab frame. A box beam was used to capture the projection of the spinel, then we used the line beam for the layer mosaicity scan with 1 $\mu$m spacing. The data were collected by a mesh scan of $\mu -\chi$ for the range of $0.8^\circ$ and $1^\circ$, with 50 and 31 steps respectively. The exposure time was set to \SI{0.5}{\sec} for each step. In total 36 mosaicity scans were collected using the same mesh scan by traversing through the 36 $\mu$m sample volume in z direction. 

\section{Results}

\subsection{Major Mineral Phase Composition}

\begin{figure}
    \centering
    \includegraphics[width=1\linewidth]{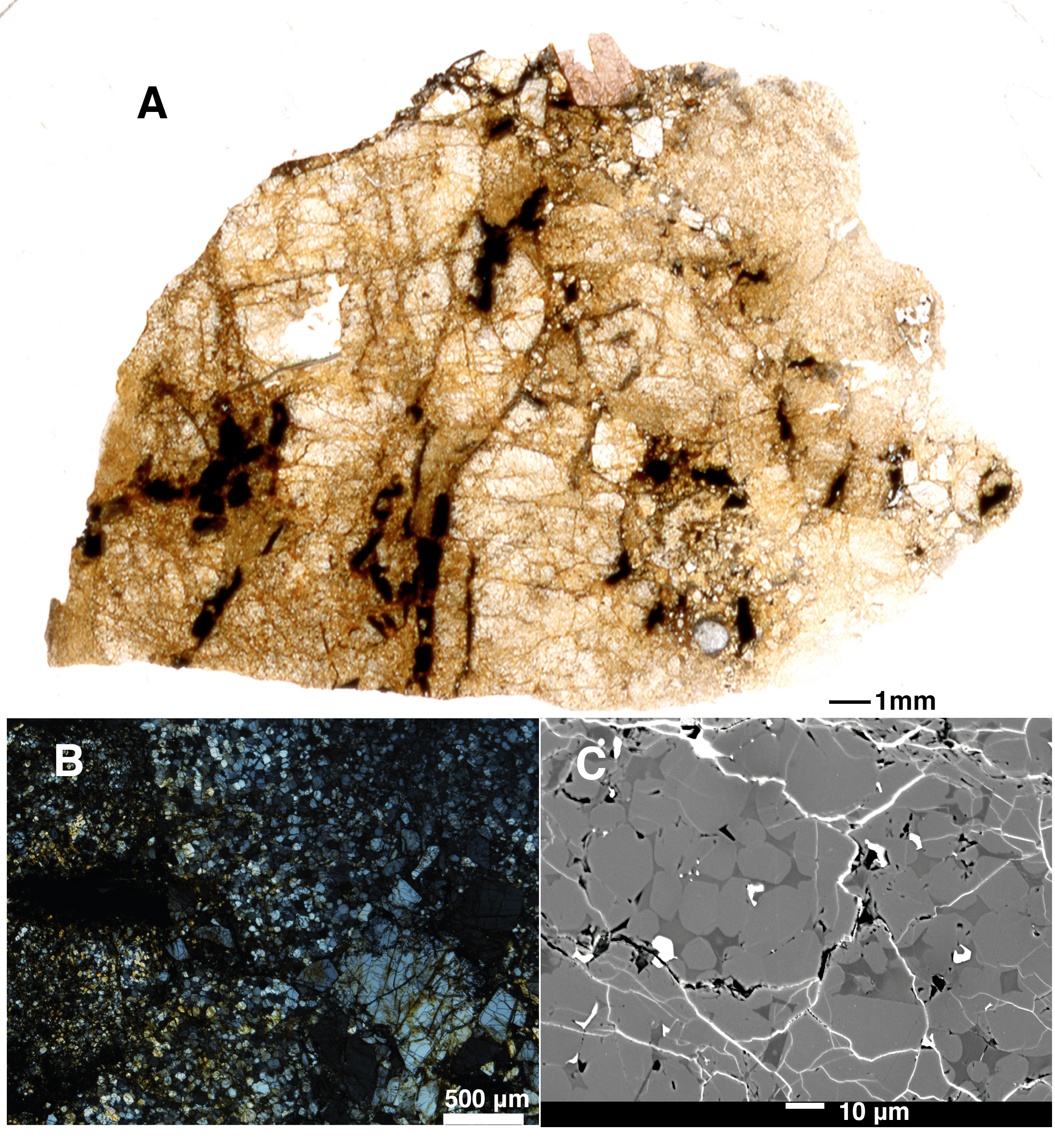}
    \caption{EET 87720 TS 57. \textbf{A} is a scanned image under Plain-Polarized Light (PPL), showing the large angular clasts with fine grained matrix. Note the prominent pink-hued Al spinel euhedral grain at top centre edge of section. \textbf{B} is a Cross-Polarized Light (XPL) image showing the zoom-in view of the fine-grained matrix that is composed of small olivine aggregates. \textbf{C} is a BSE image of olivine aggregates (lighter grey) with interstitial pyroxene (irregular shaped, darker grey).}
    \label{EET87720}
\end{figure}

EET 87720 is a polymict ureilite containing large angular clasts together with fine-grained aggregates (Fig.~\ref{EET87720}). A petrographic classification scheme for polymict ureilites was first established for DaG 319 by \citet{IkedaEtAl2000}. Based on this framework, the shock study of various ureilites by \citet{LiEtAl2021}, including EET 87720, interpreted the meteorite to be dominated by two indigenous clast types derived from the ureilite parent body: Type A coarse-grained mafic clasts and Type B fine-grained mafic lithic clasts. Morphologically, Type A clasts are characterized by grain sizes greater than 100~$\mu$m, whereas Type B clasts consist of fine-grained aggregates with grain sizes smaller than 20~$\mu$m. Here, we present detailed mineralogical and chemical characterizations of these clast populations and evaluate their possible subtypes.

\begin{figure}[!t]
    \centering
    \includegraphics[
        width=\linewidth,
        height=0.78\textheight,
        keepaspectratio
    ]{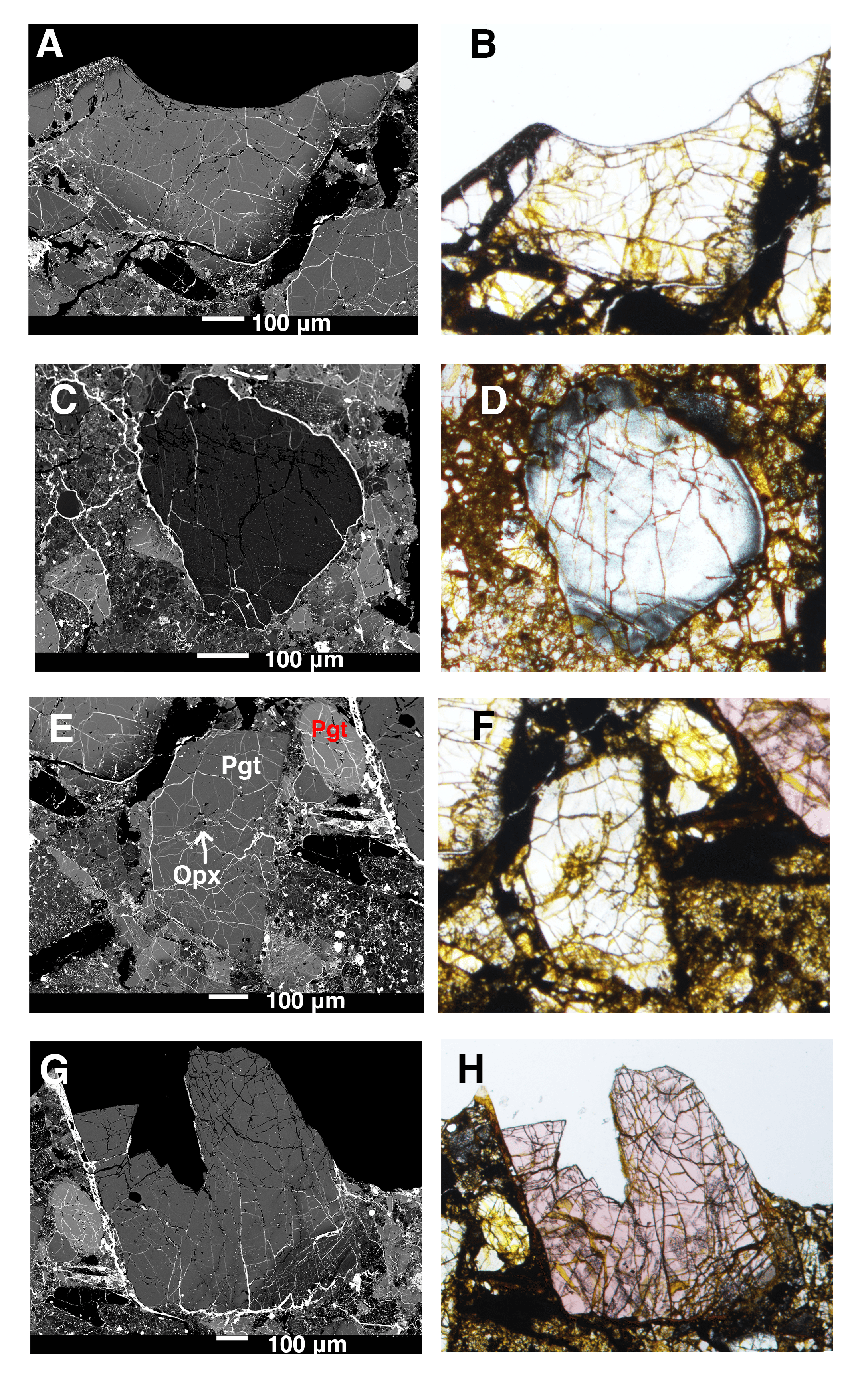}
    \caption{Example clasts in EET 87720. \textbf{A} and \textbf{B} show a BSE image and PPL image of an olivine clast, showing the Fe-reduction rim. \textbf{C} and \textbf{D} show a high-magnesian olivine clast with fine-grained opaque inclusions ($<\SI{1}{\micro\metre}$), possibly produced by an intense smelting process. \textbf{E} and \textbf{F} show a pyroxene clast, where the centre consists of a $\sim\SI{50}{\micro\metre}$ orthopyroxene enclosed within a large pigeonite ($\sim\SI{200}{\micro\metre}$). The red ``Pgt'' label marks the pigeonite that deviates from the typical ureilitic FeO/MnO trend, suggesting a cumulate origin rather than a residual origin. \textbf{G} and \textbf{H} show a euhedral spinel clast with a Fe-reduction rim at the lower right and the characteristic pink colour under plane-polarized light.}
    \label{BSE}
\end{figure}

\subsubsection{Olivine}

Olivine is the dominant mineral phase in EET 87720. Coarse angular olivine clasts (Type A; grain size $>100~\mu$m; Fig.~\ref{EET87720}) display typical ureilitic compositions with Mg\# values ranging from 74.3 to 91.8 (Table~1), and are commonly associated with pigeonite ($5.0<Wo<10.0$) or orthopyroxene ($Wo<5.0$). Based on composition, the investigated olivine clasts can be subdivided into two groups: (1) olivine clasts with Fo contents around or below 80, and (2) highly magnesian olivine with Fo $>90$.

The highly magnesian olivine clasts are texturally distinct from the less altered population and commonly exhibit reduction-related features. For example, one olivine grain exhibits a core composition of Fo$_{97.6}$ and a rim composition of Fo$_{98}$ accompanied by metallic inclusions (Fig.~\ref{EET87720}), whereas another clast shows a reduction rim of Fo$_{91.1}$ surrounding a core composition of Fo$_{79.5}$ (Fig.~\ref{EET87720}). The less altered olivine clasts yield an average composition of Fo$_{78.2 \pm 2.4}$ ($N=6$).

Fine-grained olivine aggregates (Type B; grain size $\sim10$--20~$\mu$m) exhibit a wider compositional range, with an average Mg\# of $83.2 \pm 8.7$ ($N=13$), a maximum Mg\# of 99.4, and a minimum Mg\# of 73.6.

\subsubsection{Pyroxene}

Most pyroxene phases in EET 87720, including both coarse-grained clasts and fine-grained aggregates, are characterized by low calcium contents ($Wo<10$ mol\%). Orthopyroxene ($Wo<5$ mol\%) is also present. No high-calcium pyroxene was identified in this specimen, consistent with observations reported by \citet{DownesEtAl2008} for another section of EET 87720.

The average composition of coarse pyroxene clasts is En$_{75.2 \pm 7.0}$Fs$_{18.1 \pm 5.6}$Wo$_{6.7 \pm 2.1}$. These pyroxenes occur within the common olivine--pigeonite and olivine--orthopyroxene assemblages characteristic of ureilites. The pyroxene clasts range compositionally from ferroan (Mg\# 67.5) to magnesian (Mg\# 85.2). Fine-grained pyroxene aggregates display greater compositional variability, with average compositions of En$_{81.2 \pm 12.1}$Fs$_{14.1 \pm 8.0}$Wo$_{4.7 \pm 8.0}$ and Mg\# values ranging from 77.0 to 94.0 (Table~1).

\subsubsection{Spinel}

Two coarse-grained spinel clasts, each approximately 400--500~$\mu$m in diameter, were identified in EET 87720. Both grains display a light pink color in plane-polarized light and are strongly fractured (Fig.~\ref{EET87720}A; Fig.~\ref{BSE}G--H). These spinels occur adjacent to both coarse-grained lithic clasts (Type A) and fine-grained lithic aggregates (Type B). Compositionally, both grains are highly enriched in Al, which is unusual among ureilite spinels.

Spinel compositions can generally be expressed as $A^{2+}B^{3+}_{2}O_{4}$, where the divalent $A$ site is occupied mainly by Mg$^{2+}$ and Fe$^{2+}$, whereas the trivalent $B$ site is occupied primarily by Al$^{3+}$, Cr$^{3+}$, and Fe$^{3+}$. In Spinel Grain~1, the $A$ site is dominated by Mg and Fe, with trace MnO ($\sim$0.16 wt\%). The core composition is approximately:

\[
(\mathrm{Mg}_{0.87}\mathrm{Fe}_{0.13})
(\mathrm{Al}_{1.73}\mathrm{Cr}_{0.23}\mathrm{Fe}_{0.04})O_{4}
\]

whereas the rim is comparatively more Mg-rich and Fe-poor:

\[
(\mathrm{Mg}_{0.97}\mathrm{Fe}_{0.03})
(\mathrm{Al}_{1.72}\mathrm{Cr}_{0.23}\mathrm{Fe}_{0.04})O_{4}
\]

(Table~1). Spinel Grain~1 is euhedral to subhedral and texturally distinct from the much smaller ($\sim$10--20~$\mu$m), irregular ulv\"ospinel grains reported in DaG 665 and EET 88309 by \citet{CohenEtAl2004}.

Spinel Grain~2 differs chemically from Spinel Grain~1 by containing lower Fe and Cr contents, but elevated MnO and SiO$_2$. The grain core contains $\sim$1.32 wt\% MnO and 2.7--2.8 wt\% SiO$_2$, nearly an order of magnitude greater than in Spinel Grain~1. It also contains elevated MgO ($\sim$30 wt\%), Al$_2$O$_3$ ($\sim$57 wt\%), and Cr$_2$O$_3$ ($\sim$6.5 wt\%). Assuming all Mg occupies the spinel $A$ site results in excess Mg relative to the available trivalent cations in the octahedral $B$ site. One possible explanation is that part of the MgO and SiO$_2$ occurs as an olivine-like Mg$_2$SiO$_4$ component. Under this assumption, the core composition may be approximated as:

\[
(\mathrm{Mg}_{1.04}\mathrm{Mn}_{0.029})
(\mathrm{Al}_{1.74}\mathrm{Cr}_{0.14}\mathrm{Fe}_{0.12})O_{4}
+ 0.046\,\mathrm{Mg}_{2}\mathrm{SiO}_{4}
\]

Chromium and manganese are depleted toward the rim of Spinel Grain~2. The inner rim contains $\sim$27.95 wt\% MgO, $\sim$0.79 wt\% FeO, and $\sim$69 wt\% Al$_2$O$_3$, yielding a tentative composition of:

\[
(\mathrm{Mg}_{0.93}\mathrm{Fe}_{0.017})
\mathrm{Al}_{2.04}O_{4}
+ 0.038\,\mathrm{Mg}_{2}\mathrm{SiO}_{4}
\]

The outer rim contains similar SiO$_2$ abundances but elevated Al$_2$O$_3$ (75.33 wt\%) and lower MgO (22.16 wt\%) and FeO (0.31 wt\%). Assuming excess Al$_2$O$_3$ exists outside the spinel structure, the composition may be approximated as:

\[
(\mathrm{Mg}_{0.97}\mathrm{Fe}_{0.009})
\mathrm{Al}_{2.01}O_{4}
\cdot 0.25\,\mathrm{Al}_{2}O_{3}
\cdot 0.037\,\mathrm{Mg}_{2}\mathrm{SiO}_{4}
\]

Based on the chemical analyses, the grain may alternatively reflect local non-stoichiometric behavior and unresolved submicron intergrowths.

\subsection{Oxygen Isotopes}
\begin{figure}
    \centering
    \includegraphics[width=1\linewidth]{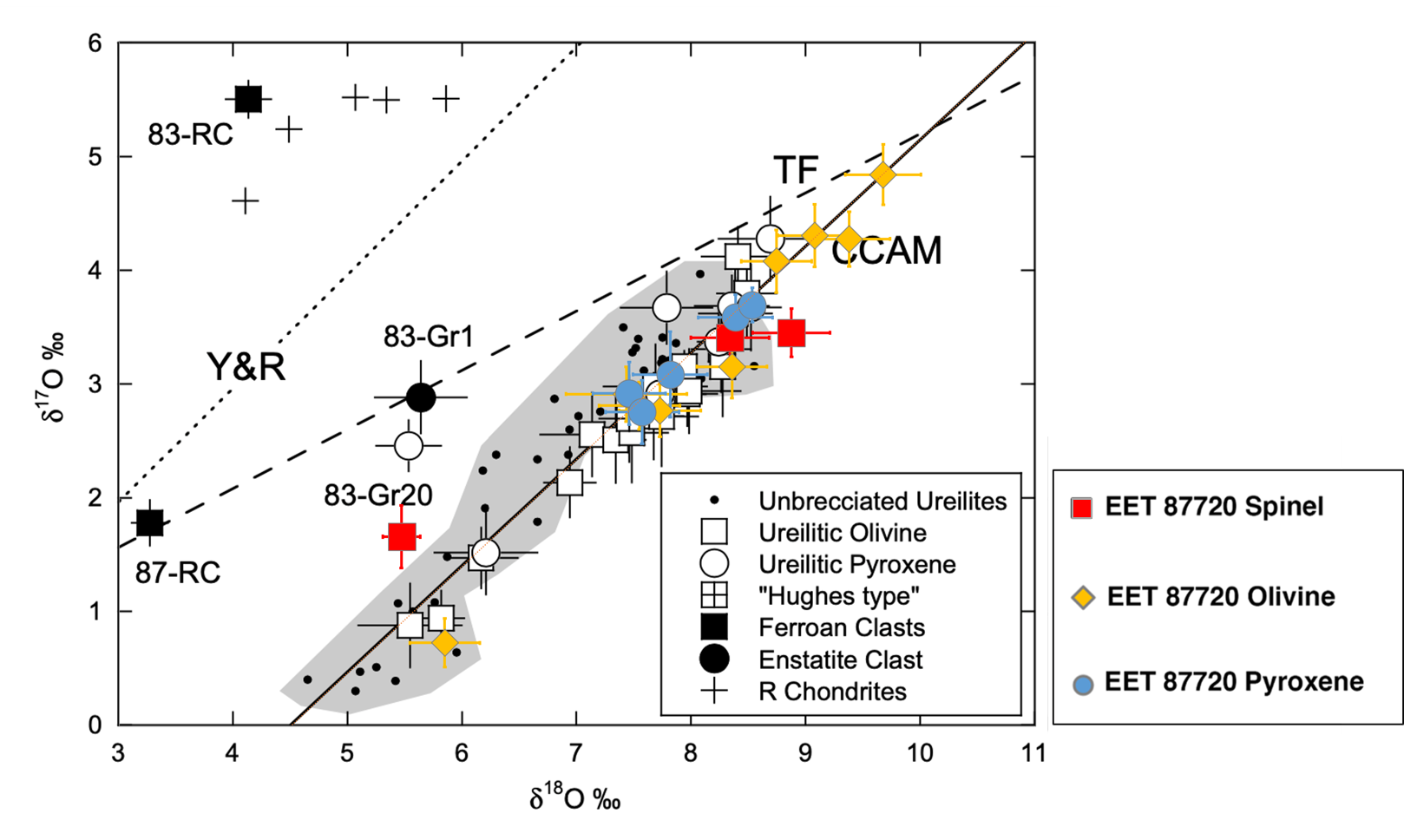}
    \caption{Oxygen isotope data from this work comparing with literature data.
    Data from this work labeled as red square for spinel, blue dots for pyroxene, and orange diamond for olivine. Figure shows oxygen isotope data for the large mafic clasts with respect to the terrestrial fractionation (TF) line, Young and Russel line, carbonaceous chondrite anhydrous mineral (CCAM) line, along with previous ureilite and other meteorite clast data from Downes et al (2008). The common ureilite oxygen isotope trend follows the CCAM line, defined by the grey region and literature date.}
    \label{OxyIso}
\end{figure}

Oxygen triple-isotope ratios of olivine and pyroxene in EET 87720 were previously reported by \citet{DownesEtAl2008} from a different specimen (TS13). However, that study did not identify Al-rich spinel grains or the highly magnesian olivine clasts documented here, with reported Mg\# values reaching only 87.6. Our investigated specimen (TS57) therefore likely contains lithic clasts that have not previously been characterized in detail, necessitating further assessment of their provenance using triple oxygen isotope analysis.

Oxygen isotope measurements in this study were obtained exclusively from large clasts ($>100~\mu$m). Multiple analytical spots (2--4) were measured on each clast to ensure representative sampling. Because polymict ureilites may contain exotic lithologies, we analyzed nine olivine clasts spanning compositions from ferroan (Mg\# 74.9) to highly magnesian (Mg\# 98.7), five pyroxene clasts ranging from Wo$_{8.9}$ to Wo$_{3.1}$, and both identified spinel clasts (Table~2).

Mean $\delta^{17}$O and $\delta^{18}$O values were calculated for each clast and plotted relative to the Carbonaceous Chondrite Anhydrous Mineral (CCAM) line and the Terrestrial Fractionation Line (TFL) defined by \citet{clayton1996oxygen} (Fig.~\ref{BSE}A). These reference lines are commonly used to constrain the origins and genetic relationships of extraterrestrial materials within the Solar System.

Most analyzed targets plot along the CCAM line (Fig.~\ref{OxyIso}). Pyroxene compositions fall within the established main-group ureilite field, whereas olivine exhibits substantially larger isotopic variability. One olivine clast, together with two additional olivine grains, plots near the upper intersection between the TFL and CCAM lines. These clasts display otherwise typical ureilitic compositions with Mg\# values between 74 and 80. Similar high $\delta^{17}$O and $\delta^{18}$O values have not previously been reported in ureilites \citep[e.g.,][]{DownesEtAl2008}. These compositions do not overlap with other known meteorite groups, and are interpreted as ureilitic in origin, extending the currently recognized oxygen isotope range of ureilites.

Highly magnesian olivine (Mg\# $>98$) plots in the lower region between the CCAM and TFL lines. This observation is consistent with the established relationship in ureilites whereby $\Delta^{17}$O decreases with increasing forsterite content \citep{DownesEtAl2008, GoodrichEtAl2004, GoodrichEtAl2015}. Oxygen isotope compositions for Spinel Grain~1 and the core of Spinel Grain~2 plot within the main-group ureilite field, whereas the rim of Spinel Grain~2 deviates from the other clasts and plots between the TFL and CCAM lines near the CM chondrite field.

\subsection{Dark Field X-ray Microscopy}
For dark-field X-ray microscopy (DFXM) analysis, we investigated Spinel Grain 1 because of the limited beamtime available and the restricted number of accessible diffraction reflections for spinel under the experimental geometry. The DFXM measurements indicate that the grain experienced significant crystal-plastic deformation.
\subsubsection{Box beam for the overview of Spinel Grain 1}
\begin{figure}
    \centering
    \includegraphics[width=1\linewidth]{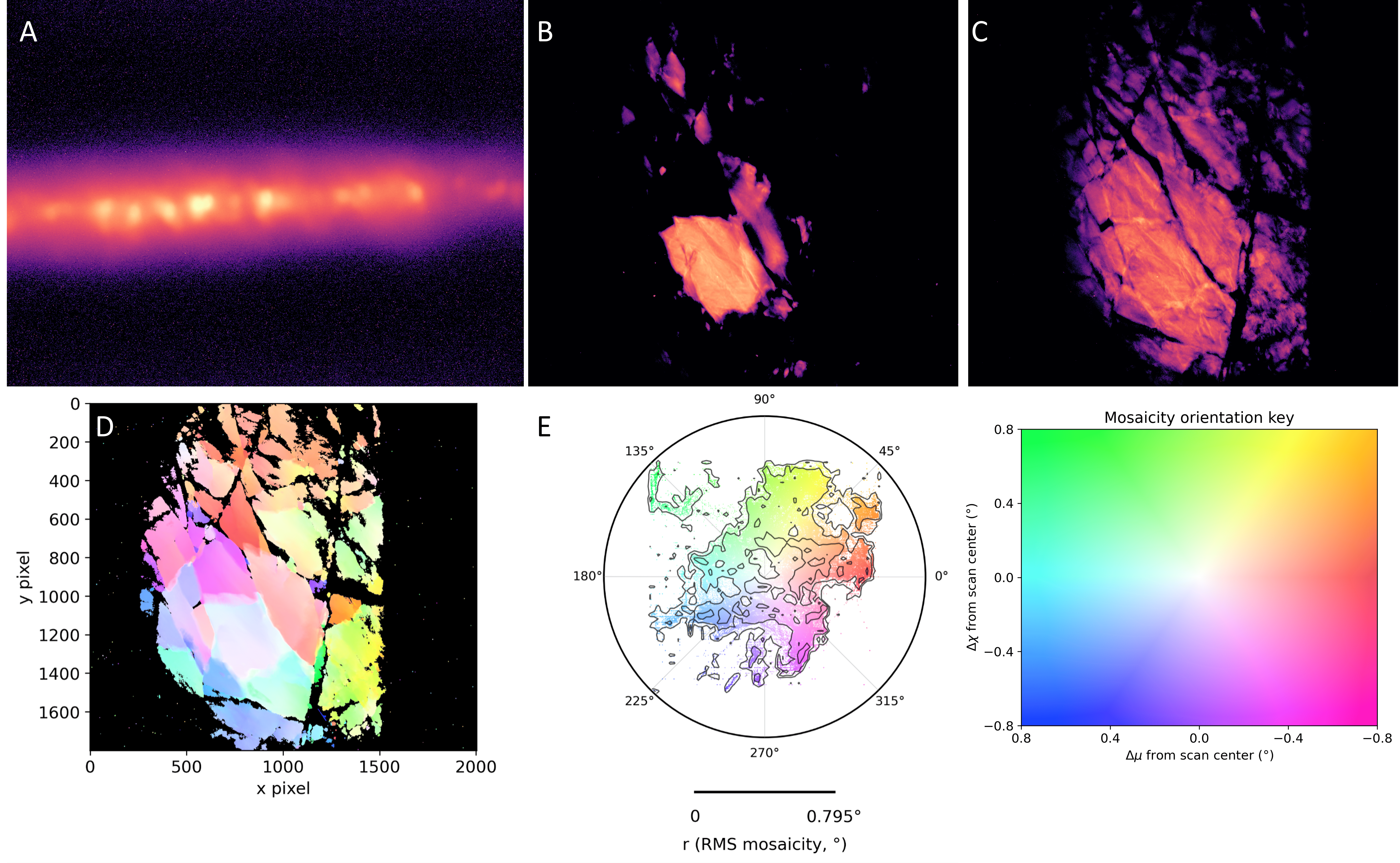}
    \caption{
    Dark-field X-ray microscopy (DFXM) characterization of Spinel Grain~1 in EET~87720.
    \textbf{A}: Reciprocal-space diffraction image recorded on the Basler diffraction camera. The diffracted intensity extends over approximately \SI{1.8}{\degree} along the $\chi$ direction, indicating a large crystallographic orientation spread within the grain.
    \textbf{B}: DFXM projection image acquired using the box beam with a $\mu$ rocking range of \SI{1.6}{\degree} at a fixed $\chi$ position of approximately \SI{0.8}{\degree}. Because of the small angular acceptance of the compound refractive lens (CRL) objective, only those regions satisfying the selected diffraction condition are imaged, resulting in only part of the grain being visible.
    \textbf{C}: Projection image reconstructed from the box-beam scan acquired over a $\mu$ range of \SI{1.6}{\degree} and a $\chi$ range of \SI{1.6}{\degree}, capturing nearly the entire grain. Pixel size is 40 nm in this field of view. 
    \textbf{D}: Reconstructed mosaicity map from the scan shown in panel~C. Colors represent the local crystallographic orientation of each reconstructed voxel. The HSV-to-RGB color key is shown on the right, and the corresponding orientation distribution is presented in panel~E.
    \textbf{E}: Orientation distribution in $\chi$--$\mu$ space calculated from the center-of-mass orientation of each reconstructed pixel. Radial distance represents the RMS mosaicity relative to the reference orientation, while color corresponds to the local orientation shown in panel~D.
}
    \label{BoxBeam}
\end{figure}

The reciprocal-space scan obtained by rocking the crystal along the $\chi$ direction reveals that the diffracted intensity extends over approximately \SI{1.8}{\degree} (Fig.~\ref{BoxBeam}A), indicating a substantial crystallographic orientation spread within the grain. A projection image collected using the compound refractive lens (CRL) objective at a single reciprocal-space position (Fig.~\ref{BoxBeam}B) captures only a portion of the grain. This reflects the limited angular acceptance of the CRL objective, whereby only regions satisfying the Bragg condition within the selected $\chi$-$\mu$ orientation are imaged. The incomplete projection therefore demonstrates that the grain contains multiple orientation variants whose angular spread exceeds the acceptance of a single DFXM image.

To capture the full orientation distribution, a box-beam scanning approach was employed. Projection images were collected while stepping through a $\chi$ range of \SI{1.6}{\degree} and rocking over a $\mu$ range of \SI{1.6}{\degree}, thereby sampling the reciprocal-space volume occupied by the diffracting crystal (Fig.~\ref{BoxBeam}C). The resulting reconstruction produces an orientation-resolved mosaicity map (Fig.~\ref{BoxBeam}D), in which color represents the local crystallographic orientation relative to the reference diffraction condition. The reconstructed grain consists of numerous subdomains separated by sharp orientation boundaries, while gradual color transitions within individual domains indicate continuous lattice rotations rather than completely discrete orientations. This microstructure suggests that the grain preserves an interconnected network of deformation-induced low-angle boundaries rather than having fragmented into randomly oriented crystallites.

To quantify the orientation distribution, the center-of-mass (COM) orientation of every reconstructed pixel was projected into $\chi$-$\mu$ orientation space to generate the orientation distribution shown in Fig.~\ref{BoxBeam}E. The reference orientation corresponds to $\chi=\SI{0.6}{\degree}$ and $\mu=\SI{1.85}{\degree}$, with radial distance representing the root-mean-square (RMS) angular deviation from this reference orientation. In this representation, $0^{\circ}$ and $180^{\circ}$ correspond to decreasing and increasing $\mu$, respectively, whereas $90^{\circ}$ and $270^{\circ}$ correspond to increasing and decreasing $\chi$. Rather than forming a uniform orientation, the orientations for each pixel group cluster into several subdomains connected by continuous subdomain boundaries. A weak preferred trend extends approximately along the $45^{\circ}$--$225^{\circ}$ direction, indicating an reversal variation between $\mu$ and $\chi$, whereby changes in one rotational component are accompanied by opposite changes in the other. Combining the $\chi$ and $\mu$ deviations for each reconstructed pixel yields a maximum Root-Mean-Squared (RMS) mosaicity of approximately \SI{0.795}{\degree}, providing a quantitative measure of the local lattice distortion within the grain for this projection scan. 

\subsubsection{Line beam for the layer scan of Spinel Grain 1}
\begin{figure}
    \centering
    \includegraphics[width=1\linewidth]{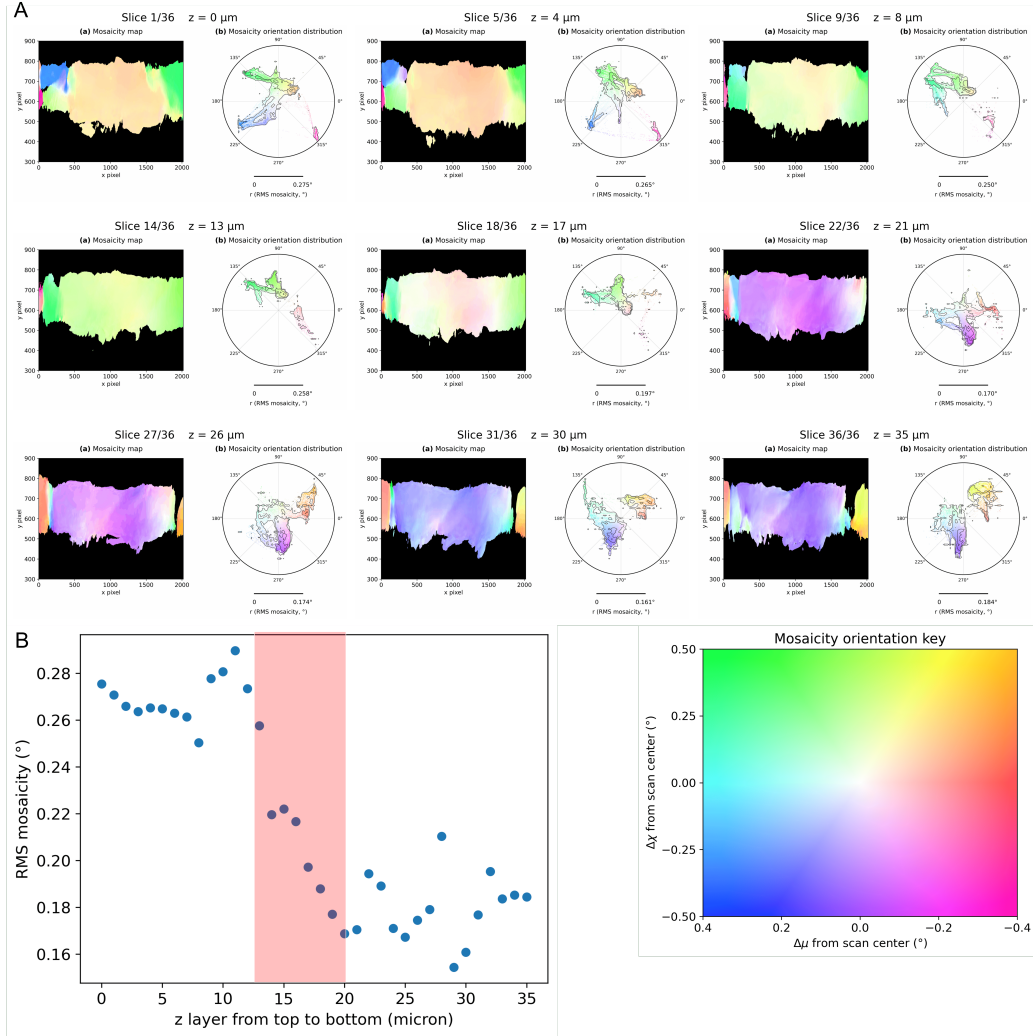}
    \caption{
     \textbf{A}: Mosaicity from line beam showing internal orientation change from top to bottom of the spinel grain. Panel (a) is the mosaicity map and Panel (b) is the Orientation distribution in $\chi$--$\mu$ space. Radial distance represents the RMS mosaicity relative to the reference orientation for each layer is shown under the panel, while color corresponds to the local orientation variation. \textbf{B}: The RMS mosaicity for 36 layers. An overall decreasing trend is observed from top layer to the bottom layer. The red box marks the region where the transition occurs, between $15\mu m$ to  $20\mu m$.  
}
    \label{Linebeam}
\end{figure}

Using a line-focused X-ray beam, a total of 36 layers were reconstructed for Spinel Grain~1 with a layer spacing of 1~$\mu$m. Layer-by-layer DFXM mosaicity reconstruction reveals that the crystal exhibits a continuously evolving internal orientation field throughout the reconstructed volume (Fig.~\ref{Linebeam}). Rather than consisting of isolated mosaic blocks, each layer contains spatially coherent subdomains separated by gradual orientation transitions, indicating that the crystallographic orientation varies continuously within the crystal.

Relative to the central reference orientation, the reconstructed layers exhibit a systematic evolution in the two orthogonal mosaicity orientation, $\chi$ and $\mu$. As the scan progresses from the upper to the lower part of the reconstructed volume, the orientation field evolves from positive $\chi$ and positive $\mu$ (Slice~1), through positive $\chi$ and negative $\mu$ (Slice~14), to negative $\chi$ and positive $\mu$ (Slice~27), before reaching negative $\chi$ and negative $\mu$ (Slice~36). This smooth progression demonstrates that the lattice orientation evolves continuously through the crystal thickness rather than by abrupt changes between discrete orientation domains.

The corresponding mosaicity orientation distributions (Fig.~\ref{Linebeam}, panel b) follow a continuous trajectory in $\chi$--$\mu$ space, spanning approximately $180^{\circ}$ around the mosaicity color wheel from Slice~1 to Slice~36. The gradual rotation of the orientation distribution confirms that the observed color variations represent systematic changes in crystallographic orientation throughout the reconstructed volume and further demonstrates that the reconstructed subdomains remain crystallographically connected in three dimensions.

The root-mean-square (RMS) mosaicity, calculated from the mosaicity orientation distribution of each reconstructed layer, exhibits a systematic decrease through the reconstructed volume (Fig.~\ref{Linebeam}, panel b \& B). Near the upper surface (0--13~$\mu$m), the RMS mosaicity remains relatively high and nearly constant at approximately 0.26--0.29$^{\circ}$. A pronounced decrease of RMS mosaicity between approximately 15 and 20~$\mu$m is observed, and then the RMS mosaicity stabilizes at lower values of approximately 0.16--0.19$^{\circ}$ towards the lower part of the crystal.

\subsubsection{Kernel Average Misorientation}
To further quantify the local crystallographic orientation gradients, Kernel Average Misorientation (KAM) maps were calculated from the center-of-mass orientation maps obtained from the $\chi$ and $\mu$ mosaicity scans. KAM measures the average local crystallographic misorientation between neighboring pixels and therefore provides a quantitative description of local lattice curvature and orientation gradients \citep{LiEtAl2026MarsStrain,ZelenikaEtAl2025}. A similar approach was previously applied to highly shocked olivine in the Martian meteorite NWA~7721 to investigate the spatial distribution of deformation structures \citep{LiEtAl2026MarsStrain}.

\begin{figure}
    \centering
    \includegraphics[width=1\linewidth]{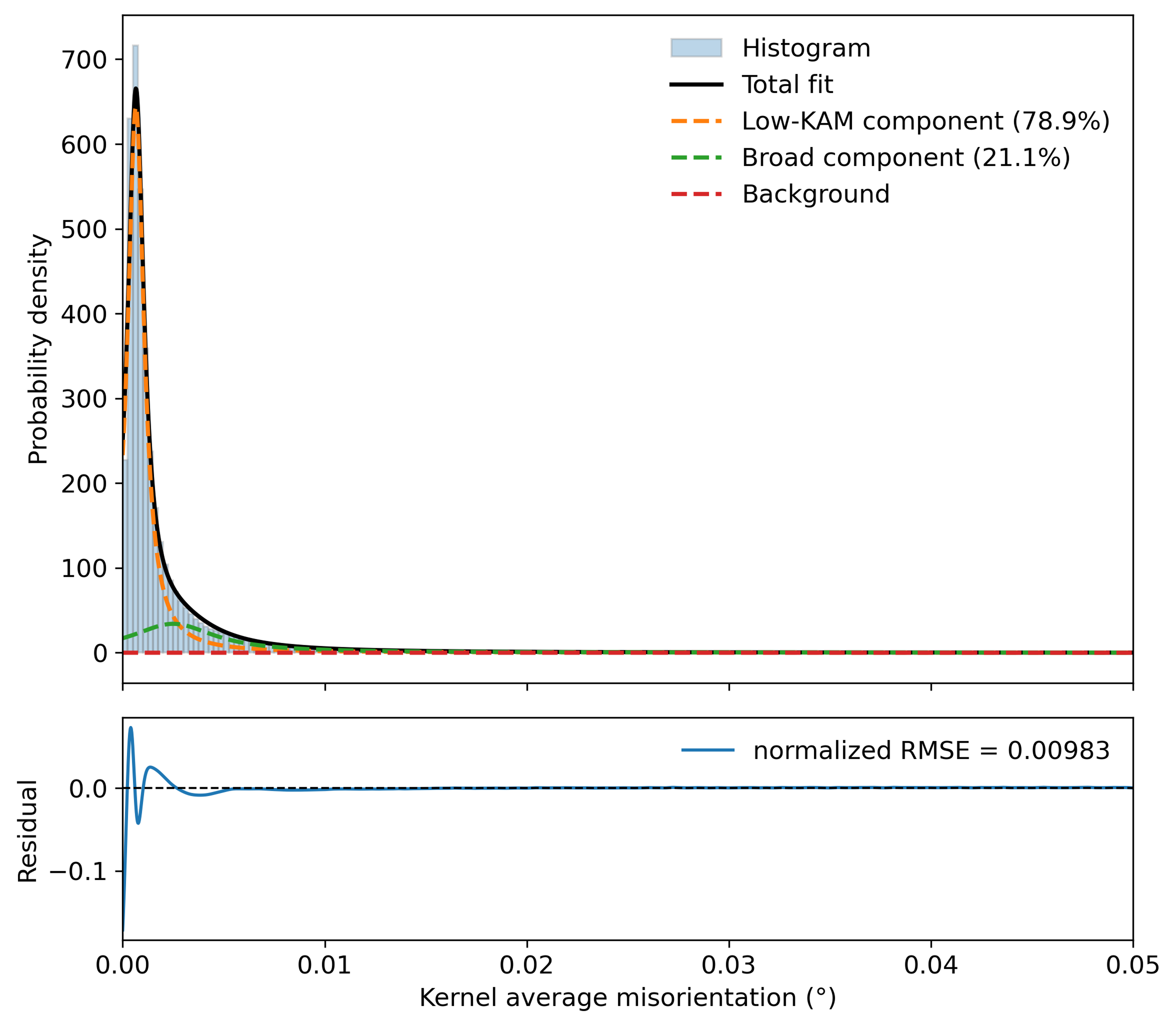}
    \caption{Peak fitting for the angle distribution of KAM in Spinel Grain 1. The data was fitted using the Best Fit for Complex Peaks (BFCP), an asymmetrical peak fitting method developed by Li et al. (2020), fitting with propotion-auto refined Pseudo-Voigt function. Two peaks were identified, where dominant small LABs contributed to the major population of the peak, and relatively larger LABs contributed to rest, causing the broadening of data peak. The goodness of fit is accessed by the normalized root-mean-squared error (RMSE) by comparing the fitting curve with the data. The value is then normalized using the total range of the peak intensity. A perfect fit would have a value of 0 for normalized-RMSE. More detail about this method can be found in Li et al's work \citep{LiEtAl2020}.     
}
    \label{KAMHisto}
\end{figure}

The KAM histogram is found to be asymmetric, consisting of a dominant low-misorientation (low KAM values) population followed by a broad tail extending towards higher KAM values (Fig.~\ref{KAMHisto}). Neither a single Gaussian nor a single Lorentzian function adequately reproduces the observed distribution. Therefore, the histogram was modeled using the Best Fit for Complex Peaks (BFCP) method developed by Li et al.~(2020), in which the distribution is represented by a weighted sum of pseudo-Voigt functions with automatically refined proportions \citep{LiEtAl2020}. The two-component pseudo-Voigt model, together with a constant background, reproduces the experimental distribution well, yielding a peak-normalized root-mean-square error (RMSE) of 0.00983 (Fig.~\ref{KAMHisto}).

The dominant component is centered at $0.00066^{\circ}$ with a FWHM of $0.0099^{\circ}$ and accounts for 78.9\% of the integrated probability density. A second, broader component is centered at $0.0025^{\circ}$ with a FWHM of $0.050^{\circ}$ and contributes the remaining 21.1\%. The first component describes the dominant low-KAM population, whereas the second component captures the extended high-KAM tail. The two weighted components intersect at approximately $0.0028^{\circ}$, which represents the transition between the dominant low-KAM population and the broader high-KAM population. This crossover value is subsequently used as a threshold for separating regions dominated by relatively low and elevated local orientation gradients.

\begin{figure}
    \centering
    \includegraphics[width=1\linewidth]{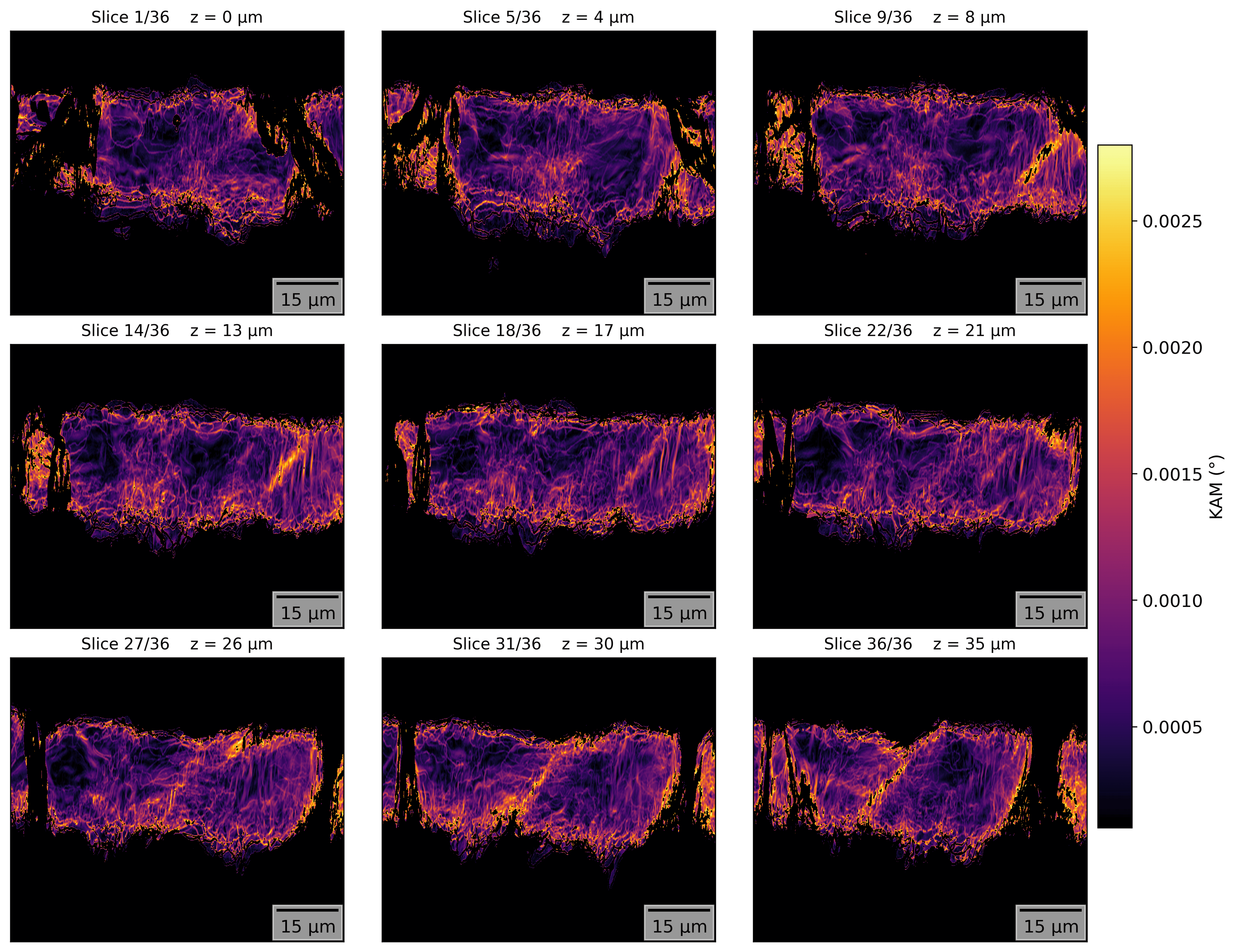}
    \caption{Spatial distribution of boundaries dominated by the high KAM ($KAM<0.0028^{\circ}$). Two types of boundaries are seen in map: the very low KAM ($0.0006^{\circ}$ ) forms the fine, irregular, and locally chaotic network while relative higher KAM  (~  $0.002^{\circ}$) forms longer and more continuous boundaries that seen successively diagonal throughout the layer.}
    \label{LowKAM}
\end{figure}

Applying the crossover threshold, regions with KAM values below $0.0028^{\circ}$ reveal a dense and spatially heterogeneous orientation-gradient network throughout the reconstructed crystal (Fig.~\ref{LowKAM}). The lowest KAM values (~ $0.0006^{\circ}$ ) form a fine, irregular, and locally chaotic network of short, curved, and branching features distributed across the interiors of the orientation domains. These weak gradients do not define sharply separated subdomains, but instead produce a pervasive background texture. They are the dominant population in the density histogram (Fig.~\ref{KAMHisto}). More organized boundaries defined by relatively larger KAM values (~ $0.002^{\circ}$ ) are seen overprinted on top of those dislocation network. These features are brighter in the KAM maps and commonly form longer, more continuous, and locally subparallel or diagonal structures that extend over the domain. These organized features are seen in the successive reconstructed layers.

\begin{figure}
    \centering
    \includegraphics[width=1\linewidth]{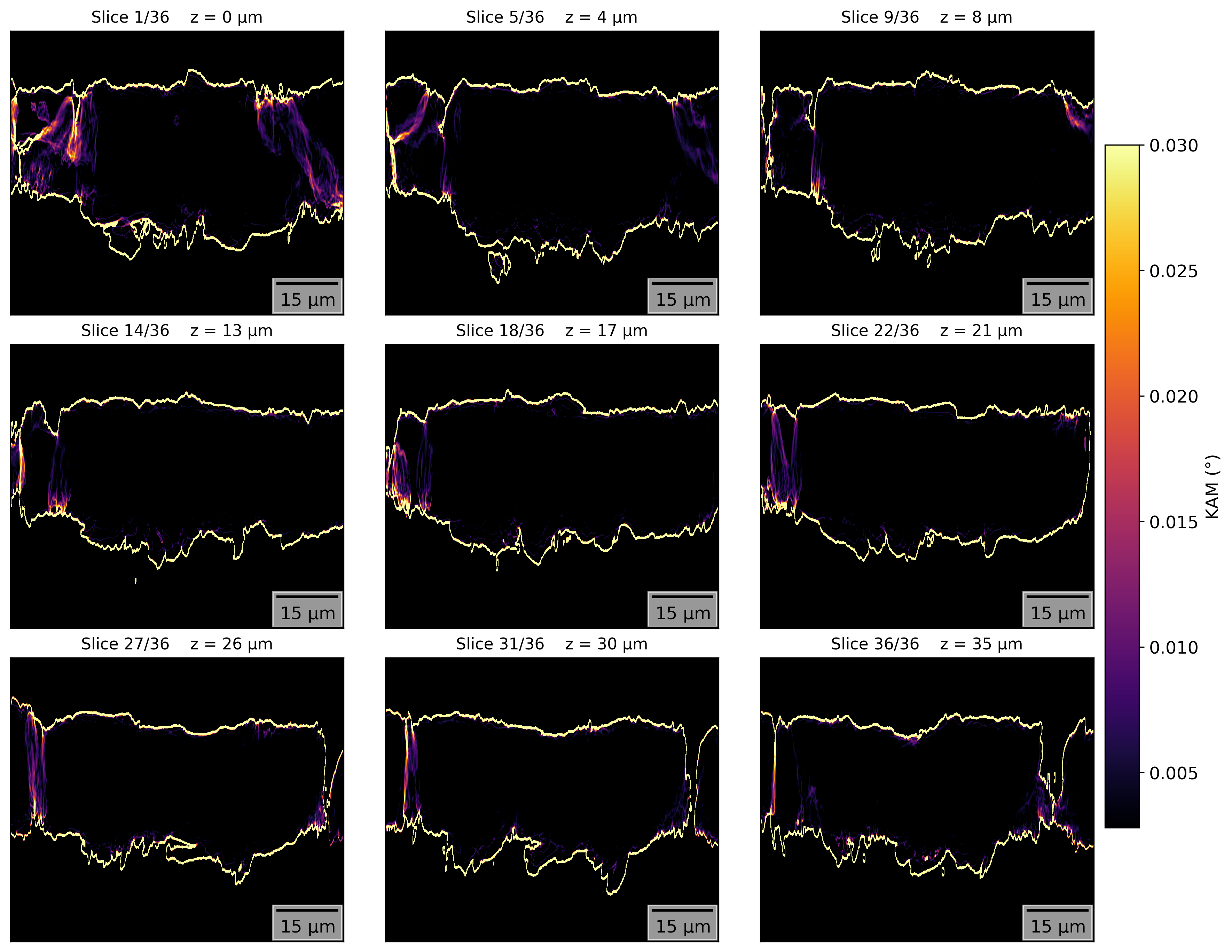}
    \caption{Spatial distribution of boundaries dominated by the high KAM ($KAM>0.0028^{\circ}$). The map highlights the margin of the scanned grain with internal subdomain boundaries. These boundaries correspond to the mosaic subdomains as seen in the mosaicity map in Fig. ~\ref{Linebeam}.}
    \label{HighKAM}
\end{figure}

Regions with KAM values greater than or equal to $0.0028^{\circ}$ substantially reveal the stronger orientation gradients within the crystal (Fig.~\ref{HighKAM}). Most subdomain interiors are removed by the threshold, thus leaving the organized boundaries of high-KAM values in the grain and near the crystal margins.  Within the crystal, moderately elevated KAM values above 0.02$^\circ$ form narrow linear and curved microstructures, some are oriented approximately subparallel to one another (e.g. Fig.~\ref{HighKAM} slice 9, 14, 27, 36). The highest KAM values (above 0.03$^\circ$) are concentrated along the external grain outline. The internal high-KAM microstructures commonly follow with transitions between differently colored regions in the mosaicity maps (Fig.~\ref{Linebeam}). 

\begin{figure}
    \centering
    \includegraphics[width=1\linewidth]{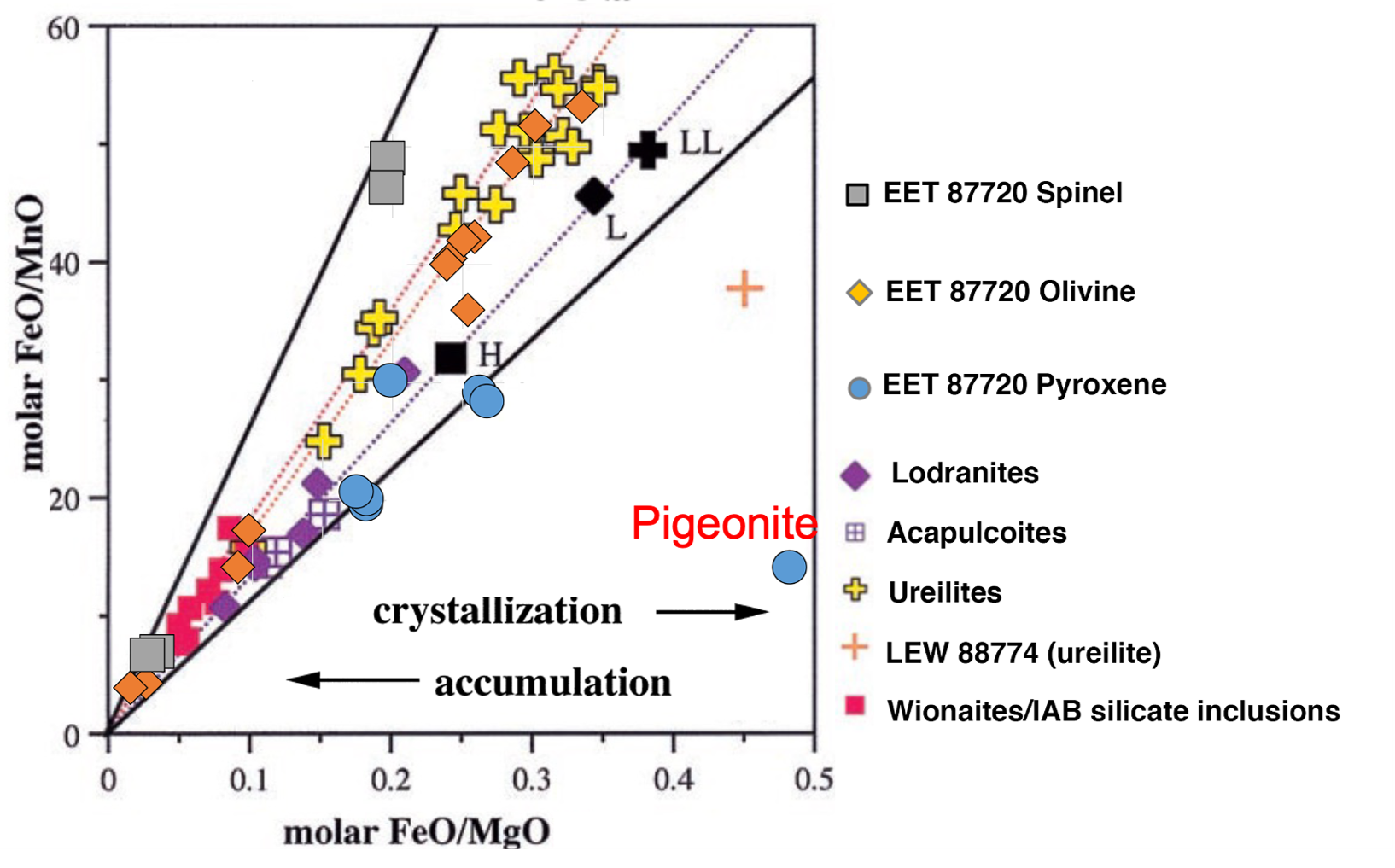}
    \caption{Fractionation of EET87720 clasts compare with literature data. Data from this work labeled as grey square for spinel, blue dots for pyroxene, and orange diamond for olivine. The molar ratio of FeO/MnO VS FeO/MgO for large mafic clasts in EET 87720 overlaying on previous work by Goodrich et al. (2004) indicating the chondritic property and accumulation-crystallization trend for differentiated rocks. Most of clasts studied in this work follows the primitive trend of ureilite, except for one pigeonite grain that shows the crystallization-accumulation trend, indicating a cumulate origin. This pigeonite grains is highlighted by the red label in Fig.~\ref{BSE}E.}
    \label{Fractionation}
\end{figure}

\subsection{Discussion}
\subsubsection{Are the clasts exotic or indigenous?}
The EET 87720 TS57 specimen studied in this work displays mineralogical similarities to specimen TS13 described by \citet{DownesEtAl2008}, in which the pyroxene is dominantly low-calcium pigeonite or orthopyroxene. However, TS57 also exhibits unique characteristics that distinguish it from previously studied ureilites.

Ureilites are known for their large variation in Mg\#, typically ranging from 74 to 92 \citep{GoodrichEtAl2004, GoodrichEtAl2015}. In contrast, EET 87720 TS57 exhibits an even broader compositional range, with both olivine and pyroxene occurring as coarse clasts and fine-grained aggregates spanning Mg\# values from 67 to 98. This unusually wide compositional range indicates overall chemical disequilibrium, most likely resulting from extensive reduction of FeO by carbon (i.e., smelting processes). Furthermore, the two newly identified spinel clasts are coarse-grained, euhedral, and enriched in Al, making them unique among currently reported ureilites. As discussed above, spinel-group oxides are extremely rare in ureilites and have previously been reported only as chromite associated with ferroan olivine (e.g., LEW 88774; \citealt{ChikamiEtAl1997, GoodrichEtAl2015}) or as ulvöspinel associated with labradoritic plagioclase (An$_{40-58}$) clasts in DaG 665 and EET 88309 \citep{CohenEtAl2004}.

Several exotic clasts were identified by \citet{DownesEtAl2008} in specimen EET 87720 TS13 based on triple oxygen isotope ratios that plot well away from the Carbonaceous Chondrite Anhydrous Mineral (CCAM) line and outside the compositional field of main-group ureilites. In contrast, the oxygen isotope ratios of all clasts analyzed in EET 87720 TS57 plot close to the CCAM line. The only exception is the outer rim of Spinel Grain~2, whose oxygen three-isotope ratios plot between the Terrestrial Fractionation Line (TFL) and the CCAM line. However, this deviation is restricted to the outer rim of Spinel Grain 2, whereas the inner core plots within the compositional range of main-group ureilites. The isotopic shift therefore likely reflects secondary modification of the grain rim rather than a distinct primary isotopic composition. The isotopically lighter rim may be related to the Fe-loss/reduction recorded in the same region due to the r. Carbon-driven reduction of FeO produces CO gas, potentially fractionating oxygen isotopes as oxygen is removed from the spinel. However, the direction and magnitude of such fractionation under ureilite conditions remain poorly constrained.

Several olivine clasts exhibit $\delta^{18}$O values greater than 9.0\textperthousand{}, exceeding the previously reported range for main-group ureilites, including ureilitic lithologies recovered from Almahata Sitta. These olivine clasts have Mg\# values between 74.9 and 79.8. Nevertheless, all analysed olivine compositions follow the Fe-loss/addition trend defined by the molar FeO/MnO versus FeO/MgO relationship (Fig.~5-3B), exhibiting chondritic Mn/Mg ratios and falling within the established ureilitic compositional field. We therefore consider these olivine clasts to be indigenous to the ureilite parent body and interpret them as extending the known oxygen isotope range of ureilites to approximately $\delta^{18}$O $\sim$ 9.7\textperthousand{}.

The two Al-rich spinel clasts identified in EET 87720 TS57 represent a previously unreported lithology within ureilites. Nevertheless, both grains possess oxygen isotope compositions that fall within the ureilitic field, with $\delta^{17}$O values ranging from 0.16 to 0.34\textperthousand{} and $\delta^{18}$O values from 5.46 to 8.87\textperthousand{}. Combined with their coarse-grained, euhedral morphology, these observations suggest that the Al-rich spinel grains are indigenous to the ureilite parent body and crystallized directly within it, rather than representing exotic condensates from the solar nebula or products of later assimilation or alteration processes.

\subsubsection{Clast formation in EET 87720 TS57}
Olivine in EET 87720 TS57 commonly exhibits Fe-depleted rims, producing reduction textures that are most likely related to a smelting process. As demonstrated above, Group~1 analyses performed on the cores of texturally unaltered olivine clasts (i.e., grains lacking inclusions or reduction textures) yield a relatively uniform composition of Mg\#78.2 $\pm$ 2.4 ($N = 6$). We interpret these grains as representing the primary olivine population that crystallized during the early evolution of the ureilite parent body. Subsequent reduction, most likely driven by carbon during smelting, removed Fe from silicate oxides (e.g., FeO), producing the reduced rims and highly magnesian olivine compositions observed in many clasts.

The Al-rich spinel clasts identified in this work also display reduced rims surrounding relatively unaltered cores, suggesting that they experienced a similar Fe-loss process to that observed in olivine. Consequently, this reduction process also resulted in a lower $\delta^{17}$O in the rim. Moreover, their occurrence raises the question regarding their possible crystallization history. The spinel grains are highly enriched in Al, yet no feldspathic minerals were identified in this specimen. Although the overall absence of feldspathic material is characteristic of ureilites, the preservation of coarse Al-rich spinel is unusual because it implactes the early environment had a locally Al-rich precursor while the early-formed spinel then escaped consumption during subsequent melt evolution. Its preservation therefore places constraints on the crystallization path of the melt, potentially reflecting conditions that favored early spinel crystallization while suppressing or delaying the formation of Al-bearing feldspathic phases.

Experimental studies have shown that pyroxene is generally incompatible with spinel under low-pressure crystallization conditions \citep{ChinnerSchairer1962, Irvine1965}. Furthermore, pyroxene grains in EET 87720 TS57 do not exhibit the Fe-depleted reduction rims observed in olivine and spinel, suggesting that they did not crystallize contemporaneousness with the spinel clasts. One pigeonite grain encloses an orthopyroxene inclusion (Fig. ~\ref{BSE}F; white label), indicating that the crystallization conditions changed during solidification of a partial melt. We therefore propose that most pyroxene crystallized during a later partial-melting event following disruption of the parent body.

Most olivine and pyroxene compositions follow the chondritic FeO/MnO versus FeO/MgO trend expected for ureilites. However, one pigeonite grain (Fig.~\ref{BSE}E\&F; red label) plots away from the main ureilitic trend (Fig. ~\ref{Fractionation}), exhibiting a composition consistent with cumulus crystallization. This observation suggests that the grain crystallized directly from an evolving melt in an igneous environment, distinguishing it from the majority of pyroxene, which is interpreted to have crystallized from the residual partial melt.

Although the partitioning behaviour of Al in olivine remains poorly constrained, we observe that Al$_2$O$_3$ concentrations generally increase with increasing Mg\#, particularly within the reduced olivine rims. This relationship suggests that Al enrichment may be associated with the same reduction process responsible for Fe loss, although the mechanism responsible for this behaviour remains uncertain.

\subsubsection{Formation of Al-rich spinel}
As described above, the spinel grains discovered in EET 87720 TS57 are unusually rich in Al. Both grains are coarse (250~$\mu$m), exhibit angular subhedral to euhedral morphologies, and preserve irregular fractures produced by later shock metamorphism. Spinel Grain~1 displays only a narrow Fe-depleted rim, whereas Spinel Grain~2 exhibits a more complex rim structure with compositional zoning characterized by variations in the relative abundances of Al and Mg. Their coarse grain size together with their euhedral to subhedral crystal morphology suggests that the spinel grains crystallized directly from a melt rather than forming through secondary alteration processes.

The formation of Al-rich spinel requires a sufficiently aluminous source. In meteorites, Al-rich spinel most commonly occurs within calcium-aluminium-rich inclusions (CAIs), where it is preserved as fine-grained aggregates interpreted to have condensed directly from the solar nebula \citep {MacPhersonEtAl2020}. Outside of CAIs, Al-rich spinel has also been reported from several lunar meteorites, including Allan Hills A81005 \citep {GrossEtAl2014, GrossTreiman2011} and Northwest Africa 11515 \citep{Li2025NWA11515}. In these samples, the preferred model involves crystallization from picritic magma that assimilated an ancient anorthositic crust, providing sufficient Al through dissolution of plagioclase-rich lithologies.

Terrestrial occurrences of Al-rich spinel provide another useful comparison. Although spinel-group minerals on Earth are dominantly Cr-rich, Al-rich spinel is characteristic of Alpine-type peridotites and mantle peridotite xenoliths entrained within basaltic lavas \citep{Irvine1965}. These rocks commonly contain magnesian olivine (Fo$_{93}$) and orthopyroxene (En$_{87-93}$), and are interpreted to originate from the upper mantle. Experimental and petrological studies suggest that crystallization of Al-rich spinel in these environments is favoured by high pressure, low oxygen fugacity, and elevated Al/Si ratios of the melt, whereas Cr-rich spinel is generally stabilized under relatively higher oxygen fugacity conditions \citep{Irvine1965}.

A mantle-like high-pressure origin appears unlikely for EET 87720 TS57. Although the size of the ureilite parent body remains debated, current estimates generally favour a relatively small asteroid. Thermodynamic modelling by \citet{WilsonEtAl2008} suggested a body with a radius of approximately 125~km, whereas \citet{GoodrichEtAl2015} proposed a radius of approximately 250~km. Under these conditions, EET 87720 is inferred to have crystallized near the parent-body surface, where lithostatic pressures would have been well below those characteristic of terrestrial upper mantle environments (likely $<10$~MPa). Consequently, crystallization of the Al-rich spinel under high-pressure mantle conditions is improbable.

Formation through assimilation of aluminous crust, as proposed for lunar meteorites, also appears unlikely. Ureilites are generally depleted in plagioclase and therefore lack an obvious Al-rich crustal reservoir capable of supplying the large amount of Al required for spinel crystallization. Likewise, direct condensation from the solar nebula is inconsistent with the textures observed in EET 87720 TS57, because nebular spinel typically occurs as fine-grained aggregates within CAIs rather than as coarse, euhedral crystals.

Assuming that the precursor materials of the ureilite parent body were compositionally similar to carbonaceous chondrites, as proposed by current models \citep{GoodrichEtAl2015}, the initial partial melt generated by short-lived radiogenic heating can be approximated by the CaO--MgO--Al$_2$O$_3$--SiO$_2$ system. Owing to its high melting temperature, Al-rich spinel would be expected to crystallize early from such a melt through the reaction

\[
\mathrm{Al_2O_3 (liq.) + MgO (liq.) \rightarrow MgAl_2O_4 (spinel)}.
\]

For spinel to be preserved, however, it must have been effectively removed from chemical equilibrium with the residual melt. Experimental studies demonstrate that spinel becomes unstable once olivine and plagioclase begin to crystallize, undergoing peritectic reactions that consume spinel during cooling \citep{Irvine1965}. Whether spinel is replaced by olivine or by plagioclase depends primarily on the MgO/CaO ratio of the melt.

Experiments by \citet{ChinnerSchairer1962} support this interpretation. Using grossular--pyrope compositions as starting materials, spinel crystallized first from the Gro$_{70}$Py$_{30}$ composition at approximately 1335$^\circ$C, followed shortly by anorthite at 1325$^\circ$C. In contrast, melts with higher MgO/CaO ratios (Gro$_{40}$Py$_{60}$) produced spinel at approximately 1470$^\circ$C, followed by forsterite at 1350$^\circ$C, with anorthite crystallizing only at lower temperatures. We suggest that a similar crystallization sequence occurred in EET 87720 TS57. An initially Mg-rich, Ca-poor melt allowed early crystallization of Al-rich spinel followed by magnesian olivine, thereby preserving spinel while suppressing extensive plagioclase crystallization. This interpretation is consistent with the overall low Ca abundance observed throughout both the clasts and the matrix of the meteorite.

\subsubsection{Spinel--olivine crystallization}
The discovery of coarse, Al-rich spinel grains in EET 87720 TS57 provides a rare opportunity to constrain the thermal conditions under which these unusual minerals crystallized. Spinel-group oxides commonly crystallize in equilibrium with olivine in mafic and ultramafic systems, and the two spinel end-members, MgAl$_2$O$_4$ and MgCr$_2$O$_4$, form a complete solid solution at magmatic temperatures \citep{Irvine1965}. Furthermore, Mg and Fe readily exchange between spinel and olivine, allowing their compositions to record crystallization temperatures through mineral-pair geothermometry. In EET 87720, both olivine and spinel display Fe-depleted reduction rims produced during the later smelting event, suggesting that they experienced a similar post-crystallization chemical evolution and may preserve evidence of an earlier equilibrium assemblage.

Several olivine--spinel geothermometers have been proposed based on Fe--Mg exchange between the two minerals \citep{Jackson1969, RoederEtAl1979, SackGhiorso1991}. However, these calibrations have been shown to produce relatively large uncertainties when applied to meteorites and other small planetary bodies \citep{KesselEtAl2007, WanEtAl2008}. Wan et al. (2008) therefore developed an alternative geothermometer based on Al partitioning between olivine and spinel, which exhibits only weak pressure dependence and has been calibrated across compositions ranging from Al-rich to Cr-rich spinels. Because the spinel grains identified in EET 87720 are exceptionally Al rich, we adopt the Al-partitioning geothermometer of \citet{WanEtAl2008}.

The crystallization temperature is given by

\begin{equation}
T(\mathrm{K})=
\frac{10000}
{0.512+0.873Y_{\mathrm{Cr}}-0.91\ln K_D},
\end{equation}

where

\[
Y_{\mathrm{Cr}}
=
\frac{\mathrm{Cr}}
{\mathrm{Cr}+\mathrm{Al}+\mathrm{Fe}^{3+}},
\]

and

\[
K_D=
\frac{\mathrm{Al_2O_3 (wt.\%,\ olivine)}}
{\mathrm{Al_2O_3 (wt.\%,\ spinel)}}.
\]

Al incorporation into olivine is considered to occur primarily through coupled substitutions involving Mg--Si exchange or the formation of cation vacancies \citep{SoulardEtAl1994, WanEtAl2008}.

A fundamental consideration when applying any mineral-pair geothermometer to EET 87720 is that the meteorite is clearly a polymict ureilite exhibiting substantial mineralogical and chemical disequilibrium. Olivine compositions span Mg\# values from approximately 67 to 98, and the meteorite records extensive reduction, shock modification, and mixing of lithologies. Consequently, it would be inappropriate to assume that all olivine grains equilibrated with the spinel prior to assembly of the breccia. Instead, our objective is to identify the subset of olivine grains most likely to preserve the original equilibrium assemblage associated with Spinel Grain~1.

To minimise the effects of later modification, only the relatively unaltered core of Spinel Grain~1 was used for the temperature calculation, as Spinel Grain~2 exhibits extensive compositional zoning and rim modification. For olivine, only Group~1 analyses, corresponding to grain cores or clasts lacking obvious reduction textures, were considered. In addition, we restricted the calculation to coarse olivine clasts that display reduction rims comparable to those developed around Spinel Grain~1, thereby selecting grains that most likely experienced a similar thermal and chemical history. Fine-grained matrix olivine was excluded because it was probably modified during post-shock thermal metamorphism \citep{Li2026Ureilite, LiEtAl2021}. Although this selection cannot demonstrate complete equilibrium, it provides the most reasonable approximation of a locally equilibrated spinel--olivine assemblage prior to subsequent reduction and impact mixing.

Using these selection criteria, the calculated spinel--olivine equilibrium temperature is $1318 \pm 43$~K ($N = 4$), corresponding to approximately $1045^{\circ}$C. This estimate agrees well with previous temperature estimates for ureilites, including approximately $1100^{\circ}$C obtained by \citet{GoodrichEtAl2004} and 1150--1300$^{\circ}$C estimated by \citet{SingletaryGrove2003} using olivine--pigeonite--melt thermometry. We therefore interpret this temperature not as the equilibration temperature of the entire meteorite, but rather as the crystallization temperature of the local spinel--olivine assemblage before subsequent reduction, shock deformation, and brecciation disrupted the original mineral associations.

The calculated temperature depends directly on Al partitioning between olivine and spinel. As shown above, olivine grains exhibiting reduction rims consistently contain higher Al$_2$O$_3$ concentrations at their rims than in their cores. Because Al substitution in olivine is accommodated through coupled substitutions involving Mg, Fe, and Si \citep{SoulardEtAl1994, WanEtAl2008}, reduction of FeO during the smelting process may promote additional Al incorporation into the olivine lattice. The removal of Fe therefore creates favourable charge-balance conditions for Al substitution, providing a plausible explanation for the systematic Al enrichment observed within the reduction rims. Consequently, the Al enrichment is interpreted as a secondary modification associated with reduction rather than a primary crystallization feature.

\subsubsection{Evidence of hierarchy of plasticity revealed by DFXM}
Experimental deformation studies on synthetic MgAl$_2$O$_4$ spinel demonstrate that crystal-plastic deformation is dominated by dislocations with Burgers vectors parallel to $\frac{a}{2}\langle110\rangle$, gliding preferentially on $\{111\}$ planes over a wide range of temperatures and strain conditions \citep{lewis1968defect,mitchell1976deformation,schafer1983shock,mitchell1999dislocations}. Transmission electron microscopy (TEM) further shows that deformation evolves from relatively straight glide dislocations at lower strain to increasingly complex three-dimensional dislocation arrangements as deformation proceeds, involving extensive dislocation interactions, cross-slip and climb \citep{mitchell1999dislocations}. These studies provide an important framework for interpreting deformation in spinel.

The DFXM reconstruction reveals three hierarchically organized populations of intracrystalline misorientation structures. The lowest-KAM population, centered near $\sim0.0006^{\circ}$, forms a pervasive irregular network that is resolved only in the KAM maps and does not correspond to discrete mosaic domains. Instead, this population most likely records weak lattice curvature or distributed orientation gradients associated with a spatially diffuse dislocation microstructure. A second population, centered near $\sim0.002^{\circ}$, defines continuous oblique slip-band-like sheets that extend through the reconstructed volume and are visible from all three laboratory viewing directions, indicating that they represent three-dimensional deformation structures rather than isolated features in individual reconstructed layers. Finally, the highest-KAM population, locally reaching approximately $\sim0.02^{\circ}$, forms narrow, continuous boundaries that separate relative orientation domains and coincide spatially with the margins of the mosaic domains observed in the orientation maps (e.g., Fig.\ref{Linebeam}). These boundaries accommodate substantially larger local lattice rotations than the surrounding deformation network and therefore represent the highest level of lattice subdivision resolved in the present reconstruction.

The coexistence of these three populations suggests that deformation was accommodated through multiple stages of lattice organization rather than through a single deformation process. The pervasive $\sim0.0006^{\circ}$ network likely represents distributed lattice curvature associated with low-amplitude dislocation interactions. The oblique $\sim0.002^{\circ}$ sheets are consistent with localized crystal-plastic deformation and resemble the planar glide structures commonly reported from experimentally deformed spinel. In contrast, the $\sim0.02^{\circ}$ boundaries possess a markedly different geometry and spatial relationship to the orientation field, defining the margins of coherent mosaic domains rather than individual slip-band-like structures. Their distinct morphology indicates that they represent a fundamentally different mode of lattice subdivision.

Because the DFXM measurements were obtained from the $(400)$ reflection, the laboratory $z$ direction is parallel to the crystallographic $[100]$ axis. Consequently, the oblique $\sim0.002^{\circ}$ sheets possibly intersect the crystallographic $a$ axis and extend throughout the reconstructed crystal volume. However, the crystallographic orientation about $[100]$ remains unconstrained, preventing unique indexing of either boundary family.

Likewise, although the $\sim0.02^{\circ}$ boundaries exhibit a distinctly different three-dimensional geometry from the oblique slip-band network, the present dataset cannot determine whether they correspond to a different crystallographic plane, a different active slip system, or a distinct mode of dislocation organization.

Although the origin of the $\sim0.02^{\circ}$ domain boundaries remains uncertain, their coincidence with the margins of the mosaic domains suggests that they accommodate lattice rotations during deformation rather than simply representing slip bands. One possibility is that these boundaries developed progressively during shock deformation as increasing dislocation density partitioned the crystal into coherent deformation domains. Alternatively, they may record the reactivation of an earlier low-angle structure during the final impact event. The present observations cannot distinguish uniquely between these scenarios.

Experimental TEM studies demonstrate that increasing deformation in spinel promotes the evolution of complex three-dimensional dislocation arrangements through interactions, cross-slip and climb \citep{mitchell1999dislocations}. The weak lattice-curvature network observed here is consistent with distributed dislocation interactions, although DFXM cannot distinguish whether these orientation gradients formed directly during shock-induced plastic deformation, developed through limited post-shock dislocation rearrangement, or partially predate the final impact event. Nevertheless, the systematic hierarchy of KAM populations demonstrates that deformation within the EET~87720 spinel was accommodated through multiple scales of crystallographic subdivision, ranging from distributed lattice curvature to localized deformation bands and coherent mosaic-domain boundaries.

Importantly, the DFXM orientation illustrates the deformation heterogeneity in this specimen. The preservation of coherent orientation domains, continuous lattice curvature, and localized three-dimensional high-KAM boundaries indicates that shock loading was accommodated through a combination of distributed crystal-plastic deformation and localized boundary formation. These structures imply heterogeneous stress transmission within the ureilite parent body, potentially enhanced by local shock heating or by disruption while the parent body remained thermally elevated. 

\begin{figure}
    \centering
    \includegraphics[width=1\linewidth]{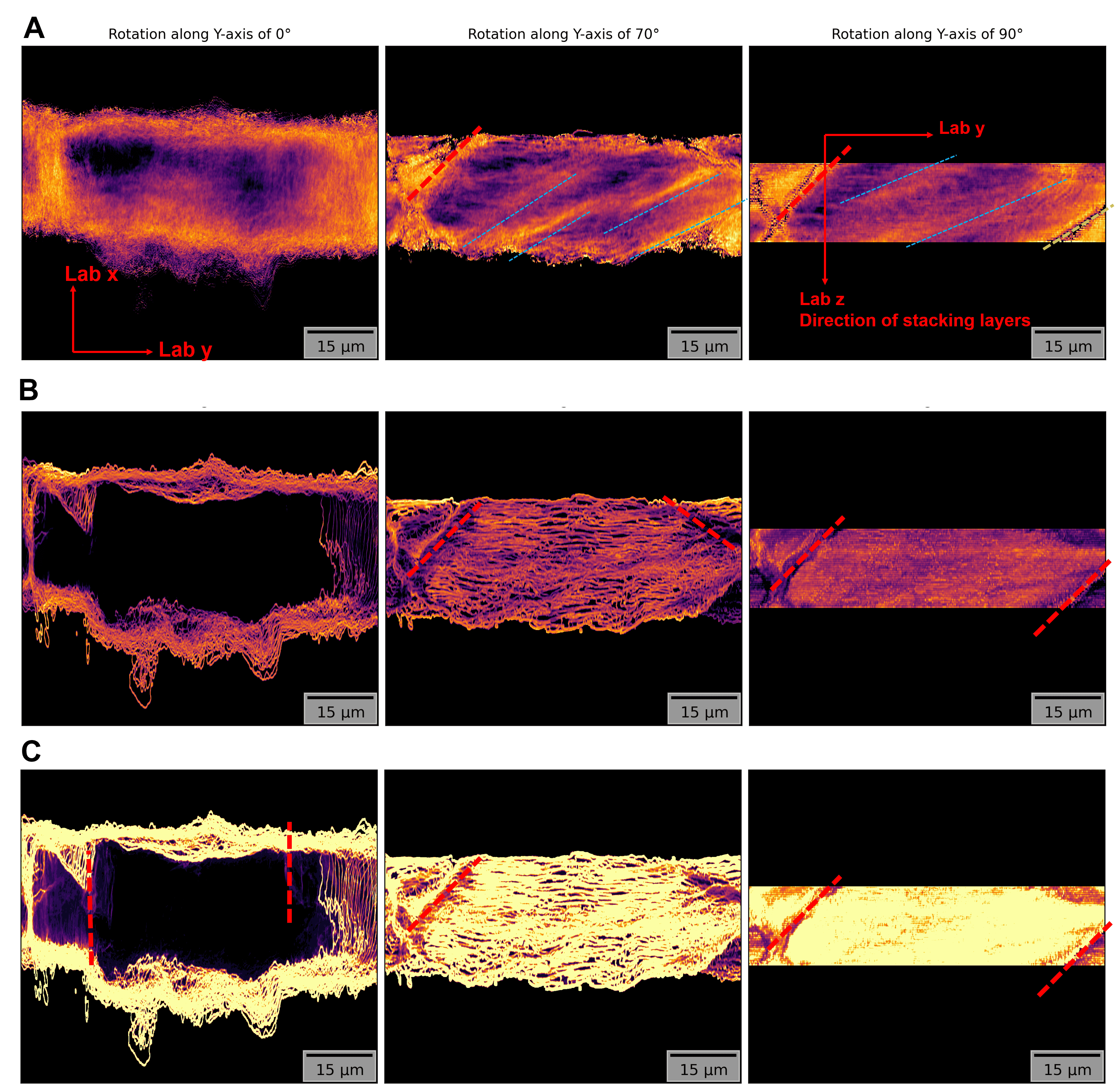}
    \caption{3D reconstruction of the 36 layer KAM rotating y-axis. Three images are select to display the slip-band features in the reconstructed volume. Fig. \ref{3DYaxis}A is the reconstruction with KAM smaller than $0.0028^{\circ}$.  Fig. \ref{3DYaxis}B is the reconstruction with KAM greater than $0.0028^{\circ}$. Blue dash lines follow the slip band trend in the figure. Red dash lines follow the different subdomain boundaries that have different orientation with low-angle KAM boundaries. Fig. \ref{3DYaxis}C is contrast enhanced volume for KAM greater than $0.0028^{\circ}$, highlighting the boundaries around 0.02$^\circ$.
    }
    \label{3DYaxis}
\end{figure}

\begin{figure}
    \centering
    \includegraphics[width=1\linewidth]{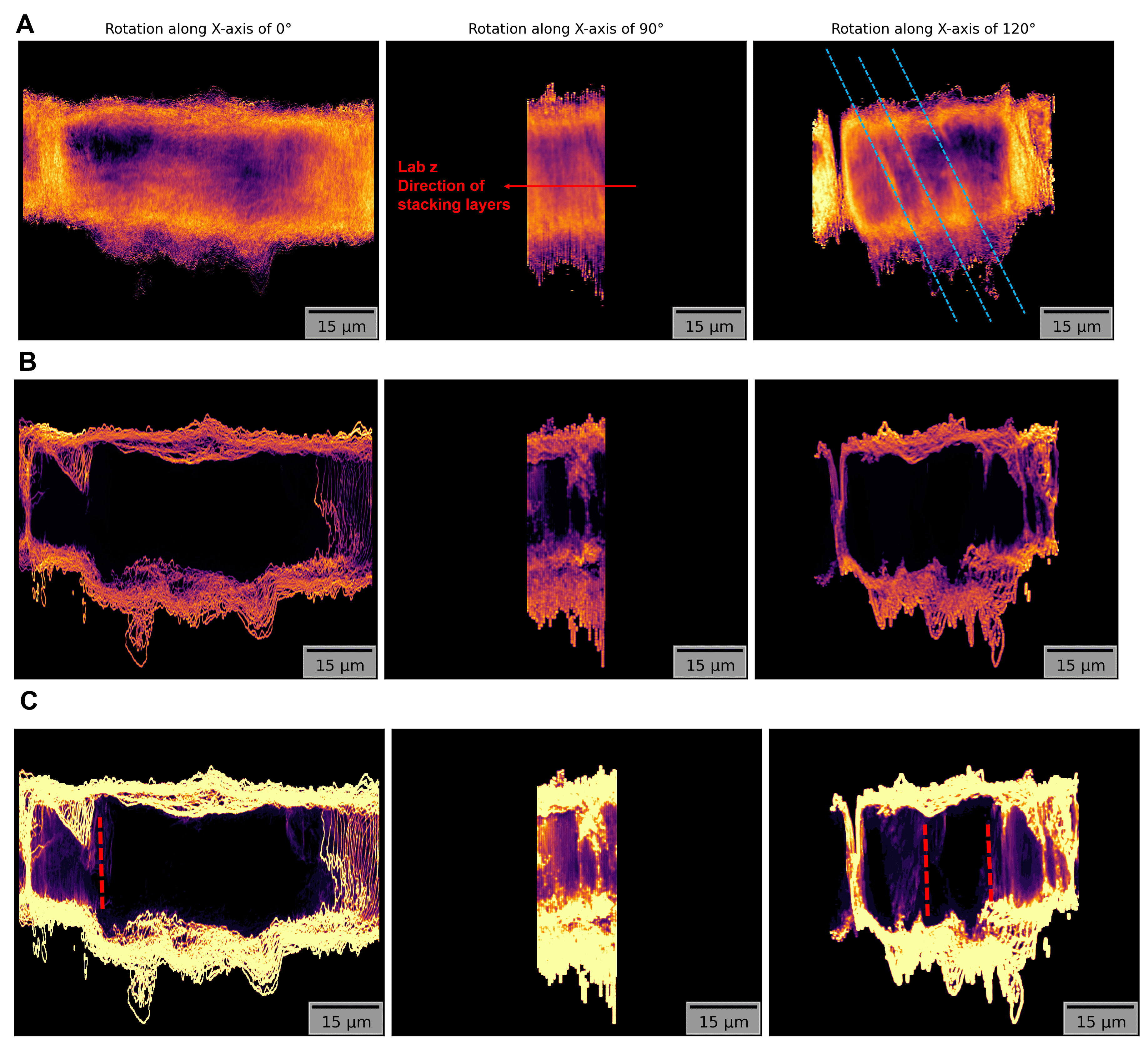}
    \caption{3D reconstruction of the 36 layer KAM rotating x-axis. Three images are selected to display the slip-band features in the reconstructed volume. Fig. \ref{3DXaxis}A is the reconstruction with KAM smaller than $0.0028^{\circ}$.  FFig. \ref{3DYaxis}B is the reconstruction with KAM greater than $0.0028^{\circ}$. Blue dash lines follow the slip band trend in the figure. Red dash lines follow the different subdomain boundaries that have different orientation with low-angle KAM boundaries. Fig. \ref{3DYaxis}C is contrast enhanced volume for KAM greater than $0.0028^{\circ}$, highlighting the boundaries around 0.02$^\circ$.}
    \label{3DXaxis}
\end{figure}

\subsection{Implications for melt evolution on the ureilite parent body}
One of the outstanding questions concerning the differentiation of the ureilite parent body (UPB) is the composition of the silicate melts generated during partial melting and their role in producing the chemically depleted ureilitic mantle. 

The occurrence of Al-rich spinel in EET 87720 provides a new constraint on the diversity of melts generated during differentiation of the ureilite parent body (UPB). 

Experimental studies suggest that the earliest low-degree partial melts of chondritic precursor were enriched in incompatible lithophile elements, including Al, Ca, Na, and Si, and the progressive extraction of these melts produced the Al-depleted olivine-pigeonite assemblages that dominate most monomict ureilites \citep{Collinet2020}. Consequently, direct mineralogical records of these early aluminous melts are uncommon. The only evidence for these phases is largely restricted to feldspathic and evolved lithologies preserved within polymict ureilites \citep{Goodrich2022}. In this context, the occurrence of a coarse Al-rich spinel in EET 87720 provides a new constraint on the diversity of crystallization products generated during early differentiation of the UPB.

As discussed above, the high Al content, Mg-rich character, low Ti abundance, and moderate Cr content of the EET 87720 spinel distinguish it from the Cr-rich spinels more commonly associated with primitive mantle assemblages. The euhedral morphology of the grain and its apparent equilibrium relationship with the surrounding silicates further suggest crystallization from a melt rather than formation solely by subsolidus replacement. Although the parental liquid cannot be reconstructed uniquely from the spinel composition alone, the stabilization of an Al-dominant spinel requires a locally aluminous chemical environment. Its occurrence is therefore consistent with crystallization from an early Al-rich melt generated during partial melting of the UPB, rather than from the strongly depleted residual mantle represented by typical ureilitic olivine--pyroxene assemblages. Considering the formation of spinel instead of plagioclase as observed in other polymict ureilites, this sample suggests an initial reservoir that produced  Al and Mg-rich but Ca-poor melt, allowing the early crystallization of Al-rich spinel followed by magnesian olivine, thereby preserving spinel while suppressing extensive plagioclase crystallization. 

When considered together with the associated olivine and pyroxene, the spinel provides complementary information on the major-element character of its parental assemblage. Olivine and pyroxene record the Mg--Fe--Ca--Si budget of the silicate system, whereas spinel preserves an otherwise poorly represented Al reservoir. The coexistence of these phases suggests that the precursor melt was sufficiently aluminous while remaining Mg-rich and poor in Ti, Ca and silica to stabilize spinel instead of plagioclase.

This interpretation is consistent with the melt predicted by the experimental model \citep{Collinet2020} involving crystallization of spinel from a localized aluminous melt generated during early partial melting, prior to the extensive extraction of incompatible-element-rich melts. The occurrence of spinel offers direct mineralogical evidence that Al-rich melt products were locally preserved during differentiation, hinting at the early presence of a silica poor ultramafic melt, Al-rich but Ca-poor. 

The pink spinels in EET 87720 may therefore preserve a crystallization product derived from a very early melt composition that has largely disappeared from the preserved ureilite record but is predicted by experimental models. It links natural mineralogy to the inferred melt evolution of the UPB. Its occurrence suggests a partial melt that is no longer represented by the dominant ureilitic lithologies, and its preservation further suggests that melt extraction was spatially heterogeneous and incomplete. Such processes allowed small volumes of aluminous material to survive subsequent reduction, impact processing, and incorporation into the polymict ureilite breccia, offering a complementary archive of the earliest stages of planetary differentiation.

\subsection{Acknowledgment}
We thank NASA/JSC for loaning Antarctic meteorite samples of EET 87720. PJAM and RLF thank the NSERC-Discovery Grants program for providing funding relating to this work. Y. Li thanks the Science International Engagement Funding (Western University) and Mitacs Canada for providing funding related to this project

\begin{table*}[htbp]
\centering
\scriptsize
\caption{Representative mineral compositions of clasts in EET 87720 determined by EPMA. Oxide abundances are reported in wt\%.}
\label{tab:mineral_comp}
\begin{adjustbox}{max width=\textwidth}
\begin{threeparttable}
\begin{tabular}{llcccccccccccc}
\toprule
Mineral & Description & SiO$_2$ & Al$_2$O$_3$ & Na$_2$O & MgO & TiO$_2$ & CaO & K$_2$O & FeO & MnO & Cr$_2$O$_3$ & NiO & Total \\
\midrule

\multirow{4}{*}{Spinel 1}
& Core & 0.10 & 57.21 & nd & 20.68 & 0.09 & nd & 0.01 & 8.25 & 0.16 & 11.37 & nd & 97.87 \\
& Core & 0.09 & 57.43 & nd & 20.90 & 0.10 & 0.01 & 0.01 & 8.32 & 0.15 & 11.37 & 0.00 & 98.39 \\
& Rim  & 0.11 & 58.57 & nd & 24.51 & 0.09 & 0.01 & nd & 2.16 & 0.21 & 11.96 & 0.02 & 97.63 \\
& Rim  & 0.06 & 59.77 & nd & 25.22 & 0.06 & nd & 0.00 & 1.29 & 0.17 & 11.89 & 0.03 & 98.49 \\

\multirow{4}{*}{Spinel 2}
& Core & 2.59 & 57.04 & 0.00 & 31.75 & 0.00 & 0.00 & 0.00 & 0.98 & 1.13 & 6.50 & 0.00 & 100.00 \\
& Core & 2.79 & 56.39 & 0.00 & 30.56 & 0.00 & 0.00 & 0.00 & 2.30 & 1.32 & 6.64 & 0.00 & 100.00 \\
& Rim  & 2.27 & 69.00 & 0.00 & 27.95 & 0.00 & 0.00 & 0.00 & 0.79 & 0.00 & 0.00 & 0.00 & 100.00 \\
& Rim  & 2.21 & 75.33 & 0.00 & 22.16 & 0.00 & 0.00 & 0.00 & 0.31 & 0.00 & 0.00 & 0.00 & 100.00 \\

\multirow{2}{*}{Olivine 1$^{+}$}
& Core & 38.89 & 0.02 & 0.00 & 40.77 & 0.04 & 0.36 & 0.00 & 18.74 & 0.44 & 0.68 & 0.02 & 99.95 \\
& Rim  & 41.08 & 0.02 & 0.01 & 49.60 & 0.00 & 0.35 & 0.00 & 8.60 & 0.49 & 0.53 & 0.01 & 100.67 \\

\multirow{2}{*}{Olivine 2$^{+}$}
& Core & 39.16 & 0.02 & 0.00 & 41.62 & 0.03 & 0.37 & 0.01 & 17.84 & 0.43 & 0.65 & 0.00 & 100.12 \\
& Core & 39.21 & 0.04 & 0.00 & 41.91 & 0.00 & 0.35 & 0.01 & 17.79 & 0.44 & 0.63 & 0.00 & 100.40 \\

Olivine 3 & N.Z. & 38.29 & 0.02 & 0.00 & 37.81 & 0.00 & 0.37 & 0.00 & 22.55 & 0.42 & 0.59 & 0.00 & 100.03 \\

Olivine 4 & N.Z. & 38.92 & 0.02 & 0.02 & 40.93 & 0.01 & 0.41 & 0.01 & 18.26 & 0.43 & 0.69 & 0.00 & 99.69 \\

Olivine 5$^{+}$ & Core & 37.51 & 0.03 & 0.01 & 39.20 & 0.00 & 0.29 & 0.00 & 19.93 & 0.40 & 0.43 & 0.00 & 97.80 \\

\multirow{2}{*}{Olivine 6$^{+}$}
& Core & 41.22 & 0.02 & 0.00 & 53.86 & 0.02 & 0.32 & 0.00 & 2.32 & 0.50 & 0.44 & 0.00 & 98.70 \\
& Rim  & 41.89 & 0.22 & 0.00 & 55.56 & 0.02 & 0.30 & 0.01 & 1.28 & 0.30 & 0.20 & 0.02 & 99.78 \\

Olivine 7 & Rim & 37.68 & 1.72 & 0.06 & 36.27 & 0.12 & 3.89 & 0.03 & 16.37 & 0.45 & 0.66 & 0.18 & 97.42 \\

Olivine 8 & Reduced & 39.71 & 0.06 & 0.01 & 48.30 & 0.00 & 0.31 & 0.00 & 7.68 & 0.53 & 0.58 & 0.02 & 97.20 \\

Olivine 9 & N.Z. & 37.82 & 0.03 & 0.00 & 38.73 & 0.00 & 0.37 & 0.00 & 20.80 & 0.40 & 0.74 & 0.00 & 98.90 \\

\multirow{2}{*}{Pyroxene 1}
& Opx incl. & 56.88 & 0.08 & 0.01 & 30.03 & 0.01 & 1.62 & 0.00 & 10.59 & 0.35 & 0.96 & 0.02 & 100.55 \\
& Pgt & 55.36 & 0.44 & 0.05 & 26.22 & 0.03 & 4.17 & 0.01 & 12.23 & 0.41 & 1.09 & 0.03 & 100.03 \\

Pyroxene 2 & Pgt & 55.41 & 0.49 & 0.04 & 26.19 & 0.06 & 4.13 & 0.00 & 12.45 & 0.43 & 1.05 & 0.01 & 100.26 \\

Pyroxene 3 & Pgt & 53.66 & 0.39 & 0.18 & 21.31 & 0.33 & 4.31 & 0.00 & 18.33 & 1.26 & 0.57 & 0.02 & 100.35 \\

\multirow{2}{*}{Pyroxene 4}
& Pgt & 55.98 & 0.76 & 0.05 & 28.73 & 0.08 & 3.44 & 0.01 & 9.25 & 0.46 & 1.11 & 0.00 & 99.87 \\
& Pgt & 56.35 & 0.77 & 0.06 & 28.97 & 0.07 & 3.42 & 0.00 & 9.34 & 0.46 & 1.09 & 0.00 & 100.51 \\

Pyroxene 5 & Opx & 56.65 & 0.42 & 0.07 & 29.84 & 0.08 & 2.55 & 0.00 & 9.23 & 0.44 & 1.07 & 0.00 & 100.35 \\

\bottomrule
\end{tabular}

\begin{tablenotes}
\footnotesize
\item Note: “N.Z.” indicates no zoning, meaning that no reduction rim or alteration texture was observed. “Reduced” indicates that the entire grain displays reduction textures attributed to smelting processes. “nd” indicates below detection limit. Superscript “+” indicates grains selected for temperature calculations.
\end{tablenotes}

\end{threeparttable}
\end{adjustbox}
\end{table*}

\begin{table}[htbp]
\centering
\scriptsize
\caption{oxygen three-isotope ratios for mineral clasts in EET 87720.}
\label{tab:oxygen_isotope}
\begin{threeparttable}
\begin{tabular}{lccccccc}
\toprule
Sample & $N$ & $\delta^{18}$O & 2SE & $\delta^{17}$O & 2SE & $\Delta^{17}$O & 2SE \\
\midrule

Spinel 1 & 6 & 8.87 & 0.34 & 3.45 & 0.21 & -1.17 & 0.13 \\
Spinel 2 (core) & 2 & 8.33 & 0.34 & 3.40 & 0.26 & -0.93 & 0.20 \\
Spinel 2 (rim) & 1 & 5.46 & 0.16 & 1.66 & 0.28 & -1.18 & 0.24 \\

Olivine 1 & 5 & 7.72 & 0.36 & 2.76 & 0.23 & -1.25 & 0.23 \\
Olivine 2 & 4 & 7.54 & 0.35 & 2.81 & 0.21 & -1.12 & 0.11 \\
Olivine 3 & 3 & 9.07 & 0.32 & 4.30 & 0.27 & -0.42 & 0.22 \\
Olivine 4 & 3 & 8.35 & 0.31 & 3.15 & 0.27 & -1.19 & 0.22 \\
Olivine 5 & 3 & 8.74 & 0.31 & 4.07 & 0.27 & -0.47 & 0.22 \\
Olivine 6 & 7 & 5.84 & 0.31 & 0.73 & 0.21 & -2.31 & 0.15 \\
Olivine 7 & 2 & 9.67 & 0.33 & 4.83 & 0.26 & -0.20 & 0.22 \\
Olivine 8 & 3 & 7.43 & 0.53 & 2.91 & 0.24 & -0.95 & 0.26 \\
Olivine 9 & 3 & 9.37 & 0.36 & 4.27 & 0.24 & -0.61 & 0.19 \\

Pyroxene 1 & 4 & 8.38 & 0.33 & 3.58 & 0.20 & -0.78 & 0.12 \\
Pyroxene 2 & 1 & 8.53 & 0.11 & 3.69 & 0.15 & -0.75 & 0.18 \\
Pyroxene 3 & 3 & 7.81 & 0.32 & 3.08 & 0.38 & -0.98 & 0.29 \\
Pyroxene 4 & 3 & 7.45 & 0.32 & 2.91 & 0.27 & -0.96 & 0.26 \\
Pyroxene 5 & 3 & 7.57 & 0.32 & 2.75 & 0.28 & -1.18 & 0.22 \\

\bottomrule
\end{tabular}

\begin{tablenotes}
\footnotesize
\item Note: $N$ represents the number of analytical targets measured on each grain. Mean values are reported when intra-grain variations were insignificant (typically within $\sim$0.1--0.2‰).
\end{tablenotes}

\end{threeparttable}
\end{table}


\FloatBarrier
\bibliographystyle{elsarticle-harv}
\bibliography{references}

@ARTICLE{ChikamiEtAl1997,
  author  = {Chikami, J. and Mikouchi, T. and Takeda, H. and Miyamoto, M.},
  title   = {Mineralogy and cooling history of the calcium-aluminum-chromium enriched ureilite, {L}ewis {C}liff 88774},
  journal = {Meteorit. Planet. Sci.},
  volume  = {32},
  year    = {1997},
  pages   = {343-348}
}

@ARTICLE{ChinnerSchairer1962,
  author  = {Chinner, G. and Schairer, J.},
  title   = {The join {C}a3{A}l2{S}i3{O}12-{M}g3{A}l2{S}i3{O}12 and its bearing on the system {CaO}-{MgO}-{Al2O3}-{SiO2} at atmospheric pressure},
  journal = {Am. J. Sci.},
  volume  = {260},
  year    = {1962},
  pages   = {611-634}
}

@ARTICLE{CohenEtAl2004,
  author  = {Cohen, B. A. and Goodrich, C. A. and Keil, K.},
  title   = {Feldspathic clast populations in polymict ureilites: Stalking the missing basalts from the ureilite parent body},
  journal = {Geochim. Cosmochim. Acta},
  volume  = {68},
  year    = {2004},
  pages   = {4249-4266}
}

@ARTICLE{DownesEtAl2008,
  author  = {Downes, H. and Mittlefehldt, D. W. and Kita, N. T. and Valley, J. W.},
  title   = {Evidence from polymict ureilite meteorites for a disrupted and re-accreted single ureilite parent asteroid gardened by several distinct impactors},
  journal = {Geochim. Cosmochim. Acta},
  volume  = {72},
  year    = {2008},
  pages   = {4825-4844}
}

@ARTICLE{GarrigaEtAl2023,
  author  = {Garriga Ferrer, J. and Rodr{\'i}guez-Lamas, R. and Payno, H. and De Nolf, W. and Cook, P. and Sol{\'e} Jover, V. A. and Yildirim, C. and Detlefs, C.},
  title   = {darfix: Data analysis for dark-field X-ray microscopy},
  journal = {J. Synchrotron Radiat.},
  volume  = {30},
  year    = {2023},
  pages   = {527}
}

@ARTICLE{Goodrich1992,
  author  = {Goodrich, C. A.},
  title   = {Ureilites: A critical review},
  journal = {Meteorit. Planet. Sci.},
  volume  = {27},
  year    = {1992},
  pages   = {327-352}
}

@ARTICLE{GoodrichEtAl2004,
  author  = {Goodrich, C. A. and Scott, E. R. and Fioretti, A. M.},
  title   = {Ureilitic breccias: Clues to the petrologic structure and impact disruption of the ureilite parent asteroid},
  journal = {Chem. Erde},
  volume  = {64},
  year    = {2004},
  pages   = {283-327}
}

@ARTICLE{GoodrichEtAl2015,
  author  = {Goodrich, C. A. and Hartmann, W. K. and O'Brien, D. P. and Weidenschilling, S. J. and Wilson, L. and Michel, P. and Jutzi, M.},
  title   = {Origin and history of ureilitic material in the solar system: The view from asteroid 2008 {TC3} and the {A}lmahata {S}itta meteorite},
  journal = {Meteorit. Planet. Sci.},
  volume  = {50},
  year    = {2015},
  pages   = {782-809}
}

@ARTICLE{GrossTreiman2011,
  author  = {Gross, J. and Treiman, A. H.},
  title   = {Unique spinel-rich lithology in lunar meteorite {ALHA} 81005: Origin and possible connection to {M3} observations of the farside highlands},
  journal = {J. Geophys. Res. Planets},
  volume  = {116},
  year    = {2011},
  pages   = {}
}

@ARTICLE{GrossEtAl2014,
  author  = {Gross, J. and Treiman, A. H. and Mercer, C. N.},
  title   = {Lunar feldspathic meteorites: Constraints on the geology of the lunar highlands and the origin of the lunar crust},
  journal = {Earth Planet. Sci. Lett.},
  volume  = {388},
  year    = {2014},
  pages   = {318-328}
}

@ARTICLE{HorstmannBischoff2014,
  author  = {Horstmann, M. and Bischoff, A.},
  title   = {The {A}lmahata {S}itta polymict breccia and the late accretion of asteroid 2008 {TC3}},
  journal = {Chem. Erde},
  volume  = {74},
  year    = {2014},
  pages   = {149-183}
}

@ARTICLE{IkedaEtAl2000,
  author  = {Ikeda, Y. and Prinz, M. and Nehru, C.},
  title   = {Lithic and mineral clasts in the {Dar} al {Gani} {(DaG)} 319 polymict ureilite},
  journal = {Antarct. Meteorite Res.},
  volume  = {13},
  year    = {2000},
  pages   = {177}
}

@ARTICLE{Irvine1965,
  author  = {Irvine, T. N.},
  title   = {Chromian spinel as a petrogenetic indicator: Part 1. Theory},
  journal = {Can. J. Earth Sci.},
  volume  = {2},
  year    = {1965},
  pages   = {648-672}
}

@ARTICLE{IsernEtAl2025,
  author  = {Isern, H. and Brochard, T. and Dufrane, T. and Brumund, P. and Papillon, E. and Scortani, D. and Hino, R. and Yildirim, C. and Rodriguez Lamas, R. and Li, Y. and Sarkis, M. and Detlefs, C.},
  title   = {The {ESRF} dark-field x-ray microscope at {ID03}},
  journal = {J. Phys. Conf. Ser.},
  volume  = {3010},
  year    = {2025},
  pages   = {012163}
}

@ARTICLE{Jackson1969,
  author  = {Jackson, E. D.},
  title   = {Chemical variation in coexisting chromite and olivine in chromitite zones of the Stillwater Complex},
  journal = {Econ. Geol. Monogr.},
  volume  = {4},
  year    = {1969},
  pages   = {41-71}
}

@ARTICLE{KesselEtAl2007,
  author  = {Kessel, R. and Beckett, J. R. and Stolper, E. M.},
  title   = {The thermal history of equilibrated ordinary chondrites and the relationship between textural maturity and temperature},
  journal = {Geochim. Cosmochim. Acta},
  volume  = {71},
  year    = {2007},
  pages   = {1855-1881}
}

@ARTICLE{LiEtAl2020,
  author  = {Li, Y. and McCausland, P. J. A. and Flemming, R. L.},
  title   = {Best Fit for Complex Peaks ({BFCP}) in Matlab for quantitative analysis of in situ {2D} X-ray diffraction data and {R}aman spectra},
  journal = {Comput. Geosci.},
  volume  = {144},
  year    = {2020},
  pages   = {104572}
}

@ARTICLE{LiEtAl2021,
  author  = {Li, Y. and McCausland, P. J. A. and Flemming, R. L.},
  title   = {Quantitative shock measurement of olivine in ureilite meteorites},
  journal = {Meteorit. Planet. Sci.},
  volume  = {56},
  year    = {2021},
  pages   = {1422-1439}
}

@ARTICLE{MacPhersonEtAl2020,
  author  = {MacPherson, G. J. and Krot, A. N. and Nagashima, K.},
  title   = {{Al}-{Mg} isotopic study of spinel-rich fine-grained {CAIs}},
  journal = {Meteorit. Planet. Sci.},
  volume  = {55},
  year    = {2020},
  pages   = {2519-2538}
}

@ARTICLE{PoulsenEtAl2017,
  author  = {Poulsen, H. F. and Jakobsen, A. C. and Simons, H. and Ahl, S. R. and Cook, P. K. and Detlefs, C.},
  title   = {X-ray diffraction microscopy based on refractive optics},
  journal = {J. Appl. Crystallogr.},
  volume  = {50},
  year    = {2017},
  pages   = {1441}
}

@ARTICLE{PoulsenEtAl2021,
  author  = {Poulsen, H. F. and Dresselhaus-Marais, L. E. and Carlsen, M. A. and Detlefs, C. and Winther, G.},
  title   = {Geometrical-optics formalism to model contrast in dark-field X-ray microscopy},
  journal = {J. Appl. Crystallogr.},
  volume  = {54},
  year    = {2021},
  pages   = {1555-1571}
}

@ARTICLE{RaiEtAl2003,
  author  = {Rai, V. K. and Murty, S. and Ott, U.},
  title   = {Noble gases in ureilites: Cosmogenic, radiogenic, and trapped components},
  journal = {Geochim. Cosmochim. Acta},
  volume  = {67},
  year    = {2003},
  pages   = {4435-4456}
}

@ARTICLE{RoederEtAl1979,
  author  = {Roeder, P. L. and Campbell, I. H. and Jamieson, H. E.},
  title   = {A re-evaluation of the olivine-spinel geothermometer},
  journal = {Contrib. Mineral. Petrol.},
  volume  = {68},
  year    = {1979},
  pages   = {325-334}
}

@ARTICLE{Rubin2006,
  author  = {Rubin, A. E.},
  title   = {Shock, post-shock annealing, and post-annealing shock in ureilites},
  journal = {Meteorit. Planet. Sci.},
  volume  = {41},
  year    = {2006},
  pages   = {125-133}
}

@ARTICLE{SackGhiorso1991,
  author  = {Sack, R. O. and Ghiorso, M. S.},
  title   = {Chromian spinels as petrogenetic indicators: Thermodynamics and petrological applications},
  journal = {Am. Mineral.},
  volume  = {76},
  year    = {1991},
  pages   = {827-847}
}

@ARTICLE{SimonsEtAl2015,
  author  = {Simons, H. and King, A. and Ludwig, W. and Detlefs, C. and Pantleon, W. and Schmidt, S. and St{\"o}hr, F. and Snigireva, I. and Snigirev, A. and Poulsen, H. F.},
  title   = {Dark-field X-ray microscopy for multiscale structural characterization},
  journal = {Nat. Commun.},
  volume  = {6},
  year    = {2015},
  pages   = {6098}
}

@ARTICLE{SingletaryGrove2003,
  author  = {Singletary, S. and Grove, T.},
  title   = {Early petrologic processes on the ureilite parent body},
  journal = {Meteorit. Planet. Sci.},
  volume  = {38},
  year    = {2003},
  pages   = {95-108}
}

@ARTICLE{SoulardEtAl1994,
  author  = {Soulard, H. and Boivin, P. and Libourel, G.},
  title   = {Liquid-forsterite-anorthite-spinel assemblage at 1 bar in the {CMAS} system: Implications for low-pressure evolution of high-Al and high-Mg magmas},
  journal = {Eur. J. Mineral.},
  volume  = {6},
  year    = {1994},
  pages   = {633-646}
}

@ARTICLE{WanEtAl2008,
  author  = {Wan, Z. and Coogan, L. A. and Canil, D.},
  title   = {Experimental calibration of aluminum partitioning between olivine and spinel as a geothermometer},
  journal = {Am. Mineral.},
  volume  = {93},
  year    = {2008},
  pages   = {1142-1147}
}

@ARTICLE{WarrenKallemeyn1989,
  author  = {Warren, P. H. and Kallemeyn, G. W.},
  title   = {Geochemistry of polymict ureilite {EET83309} and a partially-disruptive impact model for ureilite origin},
  journal = {Meteoritics},
  volume  = {24},
  year    = {1989},
  pages   = {233-246}
}

@ARTICLE{WarrenKallemeyn1992,
  author  = {Warren, P. H. and Kallemeyn, G. W.},
  title   = {Explosive volcanism and the graphite-oxygen fugacity buffer on the parent asteroid(s) of the ureilite meteorites},
  journal = {Icarus},
  volume  = {100},
  year    = {1992},
  pages   = {110-126}
}

@ARTICLE{WilsonEtAl2008,
  author  = {Wilson, L. and Goodrich, C. A. and Van Orman, J. A.},
  title   = {Thermal evolution and physics of melt extraction on the ureilite parent body},
  journal = {Geochim. Cosmochim. Acta},
  volume  = {72},
  year    = {2008},
  pages   = {6154-6176}
}

@ARTICLE{ZelenikaEtAl2024,
  author  = {Zelenika, A. and Yildirim, C. and Detlefs, C. and Rodriguez-Lamas, R. and Grumsen, F. B. and Poulsen, H. F. and Winther, G.},
  title   = {{3D} microstructural and strain evolution during the early stages of tensile deformation},
  journal = {Acta Mater.},
  volume  = {270},
  year    = {2024},
  pages   = {119838}
}

@ARTICLE{ZelenikaEtAl2025,
  author  = {Zelenika, A. and Cretton, A. A. W. and Frankus, F. and Borgi, S. and Grumsen, F. B. and Yildirim, C. and Detlefs, C. and Winther, G. and Poulsen, H. F.},
  title   = {Observing formation and evolution of dislocation cells during plastic deformation},
  journal = {Sci. Rep.},
  volume  = {15},
  year    = {2025},
  pages   = {8655}
}

@article{LiEtAl2026MarsStrain,
  author    = {Li, Yaozhu and Kal{\'a}cska, Szilvia and McCausland, Phil J A
               and Flemming, Roberta L and Hetherington, Callum J and Zhao, Bo
               and Yildirim, Can and Detlefs, Carsten},
  title     = {Remanent crustal strain on Mars in non-poikilitic olivine of {NWA} 7721},
  journal   = {npj Space Exploration},
  year      = {2026},
  volume    = {2},
  pages     = {30},
  doi       = {10.1038/s44453-026-00048-7},
  publisher = {Springer Nature}
}

@article{kita2010high,
  title={High precision {SIMS} oxygen three isotope study of chondrules in {LL3} chondrites: Role of ambient gas during chondrule formation},
  author={Kita, Noriko T and Nagahara, Hiroko and Tachibana, Shogo and Tomomura, Shin and Spicuzza, Michael J and Fournelle, John H and Valley, John W},
  journal={Geochimica et Cosmochimica Acta},
  volume={74},
  number={22},
  pages={6610--6635},
  year={2010},
  publisher={Elsevier}
}

@article{zhang2022sims,
  title={{SIMS} matrix effects in oxygen isotope analysis of olivine and pyroxene: Application to {Acfer} 094 chondrite chondrules and reconsideration of the primitive chondrule minerals ({PCM}) line},
  author={Zhang, Mingming and Fukuda, Kohei and Spicuzza, Michael J and Siron, Guillaume and Heimann, Adriana and Hammerstrom, Alex J and Kita, Noriko T and Ushikubo, Takayuki and Valley, John W},
  journal={Chemical Geology},
  volume={608},
  pages={121016},
  year={2022},
  publisher={Elsevier}
}

@article{clayton1996oxygen,
  title={Oxygen isotope studies of achondrites},
  author={Clayton, Robert N and Mayeda, Toshiko K},
  journal={Geochimica et Cosmochimica Acta},
  volume={60},
  number={11},
  pages={1999--2017},
  year={1996},
  publisher={Elsevier}
}

@article{heck2010single,
  title={A single asteroidal source for extraterrestrial Ordovician chromite grains from Sweden and China: High-precision oxygen three-isotope SIMS analysis},
  author={Heck, Philipp R and Ushikubo, Takayuki and Schmitz, Birger and Kita, Noriko T and Spicuzza, Michael J and Valley, John W},
  journal={Geochimica et Cosmochimica Acta},
  volume={74},
  number={2},
  pages={497--509},
  year={2010},
  publisher={Elsevier}
}

@article{Li2026Ureilite,
  author    = {Li, Yaozhu and McCausland, Philip J. A. and Flemming, Roberta L. and Hetherington, Callum J. and Zhao, Bowen},
  title     = {Olivine microstructure constraints on ureilite parent body deformation},
  journal   = {Journal of Geophysical Research: Planets},
  year      = {2026},
  volume    = {131},
  pages     = {e2026JE009662},
  doi       = {10.1029/2026JE009662},
  publisher = {American Geophysical Union}
}

@article{Li2025NWA11515,
  author    = {Li, Yaozhu and McCausland, Philip J. A. and Flemming, Roberta L. and Osinski, Gordon R.},
  title     = {Petrology and shock history of hybrid lunar feldspathic--troctolitic breccia Northwest Africa 11515},
  journal   = {Meteoritics \& Planetary Science},
  year      = {2025},
  volume    = {60},
  pages     = {347--370},
  doi       = {10.1111/maps.14301},
  publisher = {Wiley}
}

@article{Collinet2020,
  author  = {Collinet, M. and Grove, T. L.},
  title   = {Incremental melting in the ureilite parent body: Initial composition, melting temperatures, and melt compositions},
  journal = {Meteoritics \& Planetary Science},
  year    = {2020},
  volume  = {55},
  number  = {4},
  pages   = {832--856},
  doi     = {10.1111/maps.13471}
}

@article{Goodrich2022,
  author  = {Goodrich, C. A. and Collinet, M. and Treiman, A. and Prissel, T. C. and Patzek, M. and Ebert, S. and Jercinovic, M. J. and Bischoff, A. and Pack, A. and Barrat, J.-A. and Decker, S.},
  title   = {The first main group ureilite with primary plagioclase: A missing link in the differentiation of the ureilite parent body},
  journal = {Meteoritics \& Planetary Science},
  year    = {2022},
  volume  = {57},
  number  = {8},
  pages   = {1589--1616},
  doi     = {10.1111/maps.13889}
}

@article{mitchell1999dislocations,
  title={Dislocations and mechanical properties of {MgO}-{Al2O3} spinel single crystals},
  author={Mitchell, Terence E},
  journal={Journal of the American Ceramic Society},
  volume={82},
  number={12},
  pages={3305--3316},
  year={1999},
  publisher={Wiley Online Library}
}

@article{schafer1983shock,
  title={Shock effects in {MgAl2O4}-spinel},
  author={Sch{\"a}fer, Helmut and M{\"u}ller, Wolfgang Friedrich and Hornemann, Ulrich},
  journal={Physics and Chemistry of Minerals},
  volume={9},
  number={6},
  pages={248--252},
  year={1983},
  publisher={Springer}
}

@article{lewis1968defect,
  title={The defect structure and mechanical properties of spinel single crystals},
  author={Lewis, M-Ho},
  journal={Philosophical Magazine},
  volume={17},
  number={147},
  pages={481--499},
  year={1968},
  publisher={Taylor \& Francis}
}

@article{mitchell1976deformation,
  title={Deformation in spinel},
  author={Mitchell, T Eo and Hwang, L and Heuer, AH},
  journal={Journal of Materials Science},
  volume={11},
  number={2},
  pages={264--272},
  year={1976},
  publisher={Springer}
}
\end{document}